\documentclass[preprint]{elsarticle}

\usepackage{amsmath,amssymb,amsfonts}
\usepackage{algorithmic}
\usepackage{graphicx}
\usepackage{textcomp}
\usepackage{xcolor}
\usepackage[ruled]{algorithm2e} 
\usepackage[nolist]{acronym}
\usepackage{soul}
\usepackage{marginnote}
\usepackage{lipsum}
\usepackage[inline]{enumitem}
\usepackage{hyperref}
\usepackage{array}
\usepackage{multirow}
\usepackage{tabularx}

\newcommand{\rsec}[1]{%
  Sec.~\ref{sec:#1}%
}

\newcommand{\added}[1]{%
  {\color{blue}#1%
  }%
}

\renewcommand{\added}[1]{#1}

\newcommand{\myfigeps}[3][width=3in]{%
\begin{figure}[t]%
\centering%
\includegraphics[#1]{figures/#2}%
\caption{#3}%
\label{fig:#2}%
\end{figure}%
}
\newcommand{\myfigfulleps}[3][width=\textwidth]{%
\begin{figure*}[t]%
\centering%
\includegraphics[#1]{figures/#2}%
\caption{#3}%
\label{fig:#2}%
\end{figure*}%
}

\newcommand{\rfig}[1]{Fig.~\ref{fig:#1}}

\newcommand{\rtab}[1]{Table~\ref{tab:#1}}
\newcommand{\req}[1]{Eq.~(\ref{eq:#1})}

\newenvironment{mylist}%
{%
\begin{enumerate*}[label=(\roman*)]%
}%
{%
\end{enumerate*}%
}

\begin{document}

\title{%
Architecture and Performance Evaluation of Distributed Computation Offloading in Edge Computing
}

\author[1]{Claudio Cicconetti\corref{cor1}}%
\ead{c.cicconetti@iit.cnr.it}

\author[1]{Marco Conti}
\ead{m.conti@iit.cnr.it}

\author[1]{Andrea Passarella}
\ead{a.passarella@iit.cnr.it}

\cortext[cor1]{Corresponding author}
\address[1]{IIT, National Research Council, Pisa, Italy}

\begin{abstract}
Edge computing has been proposed to cope with the challenging
requirements of future applications, like mobile augmented reality,
since it shortens significantly the distance, hence the latency,
between the end users and the processing servers.
On the other hand, serverless computing is emerging among cloud
technologies to respond to the need of highly scalable event-driven
execution of stateless tasks.
In this paper, we first investigate the convergence of the two to
enable very low-latency execution of short-lived stateless tasks,
whose computation is offloaded from the user terminal to servers
hosted by or close to edge devices.
\added{%
We tackle in particular the research challenge of selecting the
best executor, based on real-time measurements and simple, yet
effective, prediction algorithms.
Second, we propose a performance evaluation framework specifically
designed for an accurate assessment of algorithms and protocols in
edge computing environments, where the nodes may have very heterogeneous
networking and processing capabilities.
The proposed framework relies on the use of real components on
lightweight virtualization mixed with simulated computation and is
well-suited to the analysis of several applications and network
environments.
}%
Using our framework, we evaluate our proposed architecture and
algorithms in small- and large-scale edge computing scenarios, showing
that our solution achieves similar or better delay performance than
a centralized solution, with far less network utilization.
\end{abstract}

\begin{keyword}
online job dispatching \sep serverless computing \sep computation offloading \sep edge computing \sep performance evaluation
\end{keyword}

\maketitle

\section{Introduction}\label{sec:introduction}

\added{%
Edge computing\footnote{In the scientific literature and market
press the terms ``fog'' and ``edge'' are used with overlapping vs.\
different meaning under varied circumstances, which often depend
on the specific context or application use case. The concepts
illustrated in this paper apply to a wide range of systems, including
both flavors of computing and communication systems, therefore we
do not clearly define the fog/edge terms, but rather always refer
to \textit{edge} computing to avoid ambiguity.} is generally
considered the principal enabling technology for several applications
with stringent latency requirements~\cite{7807196}.
}
With edge computing the functions that were located in a remote
data center in a \ac{MCC} architecture are called back to some point
nearer to the user~\cite{Li2017}, e.g., networking devices in the
access network with spare or added computational capabilities.
Edge computing technologies are being driven by vertical market
segments, including: \ac{IoT}, for which a reference architecture
has been published by the OpenFog
Consortium~\cite{OpenFogConsortiumArchitectureWorkingGroup2017};
the automotive and mobile network domains, which are of great
interest to the telecom industry, which has recently standardized
an inter-operable \ac{MEC} within \ac{ETSI}~\cite{Taleb2017a}.
%
%
However, edge computing has its limitations when used in a pervasive
environment with mobile devices: as shown in \rfig{intro} (a) once
an application has been provisioned on a given edge node, which is
optimal for the current user position, if the user roams towards a
different point of attachment the network must either accept increased
latency due to sub-optimal routing (top part) or pay the cost of a
migration of the application to another edge node (bottom part),
which could also cause a service interruption.

\added{%
In this paper we provide two major contributions.
Firstly, we propose to overcome the above limitation by adopting a
serverless computing approach.
Secondly, we recognize that the tools for the performance evaluation
of edge computing systems are quite limited, notwithstanding the
high interest in this technology, hence we propose a novel framework
for the performance analysis that is suitable to a wide range of
conditions of practical interest.
Both contributions are introduced briefly in the following.
}

\textbf{Serverless computing}~\cite{Varghese2018} is a novel paradigm
originating from cloud computing where clients request the execution
of short-lived ``light'' jobs, e.g, a script in a given run-time
environment, most often Python or Node.js.
Such jobs are stateless, hence they do not require a full life-cycle
management of the application nor the persistent allocation of
resources on the remote server.
This way, inherent scalability is achieved: ideally, no performance
bottleneck exists as the number of clients grows, as long
as new locations for the execution of tasks are added.
\added{%
Amazon has been among the first to offer a commercial service for
serverless computing, called AWS Lambda, closely followed by Microsoft
Azure functions and Google Cloud functions.
}
Existing solutions for serverless computing, such as Apache
OpenWhisk\footnote{\url{https://openwhisk.apache.org/}} and
Knative\footnote{\url{https://knative.dev/}}, adopt a logically
centralized load balancer to dispatch the jobs to the servers, as
show in \rfig{intro} (b).
However, such serverless computing platforms cannot be used for
computation offloading of delay-sensitive pervasive applications
in edge networks, because in this context both the clients and the
edge servers are distributed in a large geographical area and are
interconnected with links having limited capacity and introducing
significant delays, compared to the maximum tolerable latency.

\myfigfulleps%
{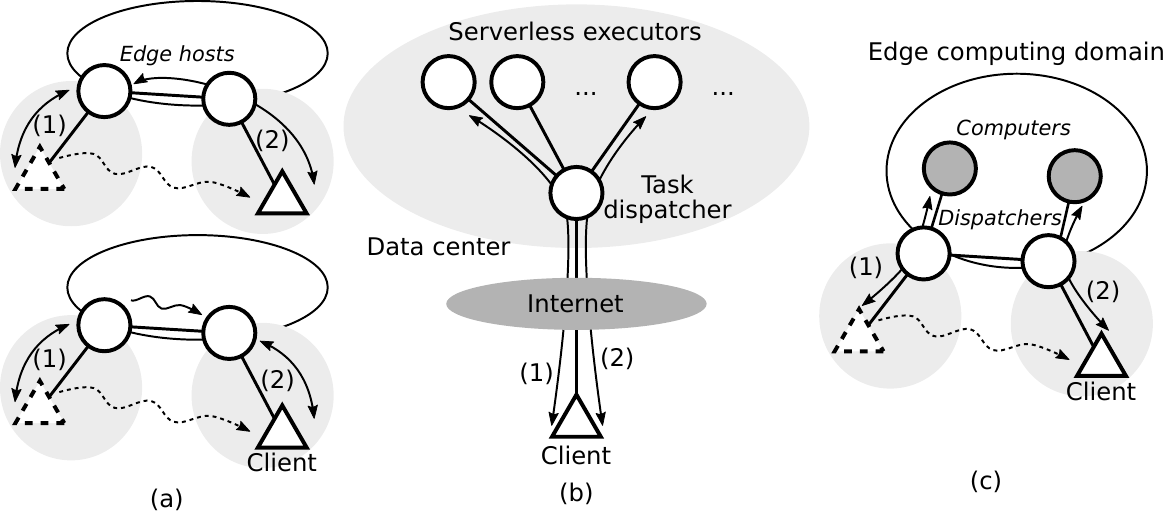}
{Comparison of architectures for (a) edge computing, (b) serverless
 computing, and (c) proposed merge of the two.}

Therefore, we propose a solution for the execution of stateless
jobs, called \textit{lambda functions}, on edge nodes as illustrated
in \rfig{intro}~(c) where we have: \textit{computers}, which have
computational capabilities to offer, because either they have spare
resources, as often found in many networking devices, or they have
been provisioned specifically for this purpose; \textit{clients},
which are \acp{UT} wishing to use said capabilities because it is
impossible or inefficient for them to perform the computation
directly; \textit{dispatchers}, which are the entry points to the
edge computing domain for clients.
Clients perform \acp{RPC} on the dispatchers, which select the most
suitable computer to run each function, forward to it the client
request, then get back the result to the client.
The \ac{RPC} is short-lived: the input is contained in the request,
the output in the response, and the call is closed immediately after
completion to avoid the persistence of long-term states in the
network.
The architecture is fully distributed: dispatchers use only their
local information to take decisions on which computer should execute
the incoming function.
Because of the ephemeral nature of lambda transactions, these
decisions are not affected by the relocation of \ac{VM}/containers
on computers, which happens far more sporadically.
%
%
The proposed architecture is a step forward compared to existing
alternatives, reviewed in~\rsec{soa:arch}.
In particular, as described in~\rsec{contrib:arch}, it achieves:
\begin{mylist}
  \item \textit{low latency}, since we cut away all the detours through
  the centralized decision point;
  \item \textit{scalability}, because the size of the decision
  problem at each dispatcher grows linearly with the number of
  computers;
  \item \textit{reliability}, as the failure of a dispatcher only
  affects the clients currently using it while the rest of the
  system runs without degradation.
\end{mylist}
%
%
%
%
Among the many research challenges associated to the proposed
architecture, in this paper we focus specifically on the \textit{online
dispatch algorithm} to select the computer that will run a given
function at a given time, which is tackled in~\rsec{contrib:dispatch}.
As discussed in further details in \rsec{soa:dispatch}, state of the art
solutions address on the one hand edge scenarios where tasks last
much longer than in our case, and therefore tasks allocations need
to be decided and changed much less frequently than in our reference
environment.
On the other hand, when tasks are ``light-weight'' as in our case,
the target scenario is typically that of tasks scheduling in
multi-core data centers, which is clearly a much more controlled
and centralized environment than ours.

\added{%
With regard to the \textbf{performance evaluation} of edge computing,
in \rsec{soa:simulation} we review the existing approaches, which
we have classified into four categories: mathematical models;
cloud models; packet-level simulations; testbed experiments.
We show that all these approaches are very well suited to capture
only a portion of the complexity incurred by an edge computing
system, where the nodes may have heterogeneous capabilities and
connectivity, and the \acp{KPI} are affected by both computation
and communication aspects, which are hard to capture at the same
time.
In~\rsec{simulation} we propose a novel framework for the evaluation
of algorithms and protocols in an edge computing environment, which
surpasses the limitations of the existing alternatives.
We argue that this framework, which relies on network emulation,
application-based virtualization, and (partial) simulation of
computation elements, is general enough to support a wide range of
environments.
To validate the framework and, at the same time, show the effectiveness
of our proposed online dispatching algorithm, we implemented the
key components of serverless edge computing and performed an
extensive campaign of experiments, whose results are reported
in~\rsec{eval}.
The online dispatch algorithm has been compared with a centralized
approach, as commonly found in serverless computing in data centers,
and a known online algorithm from the literature~\cite{Tan2017}.
}

\section{State of the art}\label{sec:soa}


\added{%
In this section we review the state of the art in the literature,
separately for the main contributions of this work: architecture
for the convergence of the edge computing and serverless computing
paradigms (\rsec{soa:arch}); distributed dispatching of lambda
functions to computers (\rsec{soa:dispatch}); modeling and large-scale
simulation of edge computing environments (\rsec{soa:simulation}).
}

\subsection{Converged serverless edge computing architecture}\label{sec:soa:arch}

In the scientific literature there are several proposals on how to
realize computation offloading in edge computing.
However, the vast majority are based on some form of lightweight
orchestration, as compared to having a true ``cloud'' with \acp{VM},
by scaling down cloud-oriented paradigms to less powerful servers
and faster dynamics.
Examples include Picasso~\cite{Lertsinsrubtavee2017}, from which
we reuse the concept of providing the applications with an \ac{API}
whose routines are executed by the network in a manner transparent
to clients, and \textit{foglets}~\cite{Saurez2016}, which use
containers for an easier and faster migration of functions based
on situation-awareness schemes.
Both studies put forward efficient ways to periodically tune the
deployment of containers in edge servers, which is complementary
to our work, where we focus instead on the short time scale dispatching
of tasks once a given deployment in place.

In addition to generic architectures, such as what we propose in
this paper, there are solutions tailored to specific scenarios.
In~\cite{Krol2017} the authors exploit \ac{ICN} to realize a paradigm
called \textit{Named Function as a Service (NFaaS)}, where the
functions are automatically distributed over \ac{ICN}-enabled servers
based on their utilization.
Again, this could be a suitable complement to our present contribution,
where we focus mostly on the short-term dispatch problem over an
interval small enough that we can assume that the functions on the
computers are stable.
It is interesting to point out that to reduce the latency of setting
up/tearing down the application, it is suggested that the servers
employ \textit{unikernels}~\cite{Madhavapeddy:2013:URV:2557963.2566628},
which are an extreme form of containerization.
\added{%
In the context of \ac{IoT}, a cloud-edge architecture is proposed
in~\cite{Filip2018}, which employs micro-services to run computation
on heterogeneous edge or cloud nodes: different algorithms are
proposed to schedule tasks from the clients, in the assumption that
a centralized scheduling engine can estimate at fine grain both the
task processing time and the time it will be put into service, which
is advocated to be realistic in some conditions, i.e., with \textit{a
priori} knowledge on the application algorithms and continuous
monitoring of the computation engines.
When applying the constraints of our use case to the scheduling
policies considered there, they basically collapse into scheduling
every task to the computer with shortest expected processing time,
which is exactly what we propose in~\rsec{contrib:dispatch}.
}
Finally, also in the context of \ac{IoT} in~\cite{Singh2017a} the
authors propose an architecture to distribute jobs to a set of
gateways by means of dynamic data plane manipulation: since all the
client requests pass through the \ac{SDN} controller, the latter
can estimate the arrival process and allocate new requests to servers
accordingly.
Unfortunately, such information is not available in a fully distributed
approach such as ours.

\subsection{Distributed dispatching of lambda functions}\label{sec:soa:dispatch}

From a high level perspective, the distributed dispatching problem
can be described as follows, from the point of view of a given
dispatcher.
There are a number of lambda functions (or jobs) that will arrive
over time and will have to be dispatched to a pool of computers,
with the goal of minimizing their response time.
The arrival process and the execution times are not known \textit{a
priori}.
So far, this is a set-up for a classical multi-server scheduling
problem, that has been extensively studied in the literature due
to its huge importance in designing efficient schedulers in multi-core
systems, both stand-alone and in grid/cloud environments.
A known result is that no online algorithm\footnote{An \textit{online}
algorithm is one that takes a decision on a per-job basis and is
not allowed to remain idle, in contrast to \textit{offline} algorithms
that have knowledge of past and future task arrivals, hence, may
decide to delay a task even when the servers are idle to maximize
the objective function.} can have a bounded competitive ratio, see,
e.g.,~\cite{Anand2012}.
In the same work the authors also propose a method to derive
approximation algorithms that have a bounded competitive ratio in
a speed augmentation model, i.e., by assuming that the online
algorithm is given extra resources.
One major difference with an edge computing scenario is that the
multi-server scheduling problem assumes that the servers are used
exclusively and that the policy for the execution of the jobs within
each server is also under control.
Both assumptions are false in our system:
\begin{mylist}
  \item any computer can be assigned jobs by multiple non-communicating
  dispatchers, and
  \item we cannot reasonably assume to have influence (or even
  insight!) on the scheduling within computers, which are highly
  heterogeneous (ranging from a \ac{WLAN} routers to multi-core
  servers in a \ac{MNO} core network) and shared (e.g., a telco
  server may offer computational capabilities while also implementing
  \acp{VNF}).
\end{mylist}
Furthermore, different computers may have different communication
latencies with the dispatcher, and they may vary over time since
the network of an edge computing domain is expected to be also
(well, actually mostly) used for Internet access.

A closer view is taken in~\cite{Tan2017}, which in fact deals more
specifically with edge computing.
The authors propose an approximation algorithm to minimize the total
weighted latency of the jobs, where the weight is assumed to be
generically related to the delay-sensitiveness of the job.
The algorithm is proved to be $\mathcal{O}(1/\epsilon)$-competitive
in the $(1+\epsilon)$-augmented problem.
The algorithm takes into account the communication latencies, which
are assumed to be known for a given job, and it requires the
processing time of every incoming job if executed on any given
computer, which in general is not available.
In our proof-of-concept we implemented the algorithm proposed
in~\cite{Tan2017} when using emulated computers, which can provide
the exact processing time of a job if no other arrives until its
completion, as described in \rsec{simulation}.
Comparison results with our proposed algorithm are reported in
\rsec{eval:large}.
\added{%
The authors in~\cite{Sthapit2018} also tackle specifically edge
computing by defining centralized and distributed algorithms to
solve an optimization problem: jointly optimize the processing time
and energy consumption at mobile users, where the network is modeled
as a network of queues.
In the distributed version, the authors study the trade-off
between the communication overhead for synchronization and the
performance of the algorithm.
In this work, since we focus on pervasive applications and the edge
nodes are assumed to have limited capabilities, we consider only
the limit case where there is \textit{no synchronization at all}
between the edge nodes, and show through emulation experiments that
this approach is not detrimental compared to the case where there
is a single entity making network-wide choices.
}

\added{%
Some other works explored the concept of ``tasklets'', which are
similar to micro-services, because they are invoked by clients as
functions, but are intended to be used to realize in-network
computation with user devices.
}
More specifically, Edinger \textit{et al.}~\cite{Edinger2017}
envision a system where computation consumers ($\simeq$~\textit{clients})
contact brokers ($\simeq$~\textit{dispatchers}) that direct them
to the most suitable computation producers ($\simeq$~\textit{computers}),
who are then contacted directly for the execution of tasklets.
In the work above the most important scientific result is a scheduling
algorithm that reduces the number of execution failures by estimating
the reliability of producers.
\added{%
The problem of task partitioning is studied in~\cite{Schaefer2018},
along with solutions to reactively/proactively migrate the \acp{VM}
to minimize the overall task execution time despite failure of
devices offering computation.
}
Such contributions are not directly applicable to our case because
the computers are devices specifically committed to computation
offloading and, thus, can safely be assumed to disconnect or fail
very sporadically.
Furthermore, the tasklet scenario is flat, and brokers are introduced
merely for scalability reasons, whereas an edge computing network
is well structured, with clients logically separated from computers
by access network gateways, which we use as dispatchers: in this
structure the latter can easily monitor execution of lambda requests,
which always pass through them, unlike brokers for tasklets.

\added{%
\subsection{Edge computing modeling and simulation}\label{sec:soa:simulation}

The performance evaluation of serverless in edge computing is a
scientific challenge \textit{per se}.
We can distinguish four main categories in the literature.

We consider the pure mathematical approaches first, where a vast
number of them is aimed at determining the best allocation of
\acp{VM}/containers under various constraints.
A fairly comprehensive and recent model suited to placing Docker
containers on edge nodes is~\cite{Smet2018}, which uses an \ac{ILP}
problem formulation.
Modeling the edge network as a network of queues is also found in
some works, e.g.~\cite{Sthapit2018}.
While the use of a simplified model makes the problem tractable in
mathematical terms, this inevitably hides some aspects, which in
the case of devices with limited computation and communication
capabilities are crucial to capturing key aspects of the system
dynamics, such as the overhead of resource sharing, communication
protocol latencies, and processing delays.

Some of these aspects are modeled accurately in the second approach,
i.e., cloud-oriented simulator, reviewed more extensively
in~\cite{Al-Dhuraibi2018}.
The most widely used such simulator is
CloudSim~\cite{Calheiros:2011:CTM:1951445.1951450}, which has been
extended to also support specifically, e.g.,
containers~\cite{doi:10.1002/spe.2422} and edge computing
scenarios~\cite{Gupta2017}.
These simulators are aimed at evaluating the relative performance
of allocation algorithms, both off-line and dynamic, also introducing
a basic simulation of the effects due to networking, especially in
the edge-oriented flavors.
A key advantage over the pure mathematical approaches is that they
can model a wide variety of workloads, also taken from real-life
datasets, and still support the performance evaluation of large-scale
scenarios at very reasonable computational cost (see,
e.g.,~\cite{FERNANDEZCERERO2018160} for a highly-scalable simulation
architecture written in Scala).
However, to the best of our knowledge, it is not possible with any
of them to assess the performance of real applications or system
components, because the simulation fully happens within its own
realm, with no exchange points with the real world.
For the sake of completeness, we mention that the evaluation of
real applications using a CloudSim derivative has been
attempted~\cite{doi:10.1002/spe.2124}, but with severe limitations
and achieving a level of maturity inferior to the main tool.

This limitation is overcome by using discrete-event packet-level
simulators, which most often allow real packets to be injected into
the engine, and at the same time offer very detailed model of the
physical and networking environment.
Examples include Omnet++~\cite{Qayyum2018} and ns-3~\cite{Zhou2019}.
Unfortunately, real-time emulation requires that the simulation
engine is fast enough to process all the events, which limits
significantly the size of the scenario that can be evaluated.

Finally, the fourth approach is the execution of experiments
in a real edge/cloud environment.
This direction has been followed to compare existing serverless
frameworks (open source in~\cite{Mohanty2018}, commercial
in~\cite{Lynn2017}).
While this provides obviously an environment as close as possible
to a production system, there are several disadvantages: experiments
tend to be extremely expensive in terms of both time and equipment;
it is difficult to reproduce experiments because of the many factors
affecting performance in real life, especially if the platforms are
not fully owned by the experimenter; the scale of the experiments
is limited by the physical resources owned or accessible.

\begin{table}[htbp]%
\caption{V}%
\begin{center}%
{\small%
\include{tables/}%
}%
\label{tab:}%
\end{center}%
\end{table}%
C
{sim-comp}%
{Qualitative comparison of different approaches to performance
 evaluation of edge computing systems.}

To overcome the limitations introduced by each respective approach,
while relinquishing only a fraction of their advantages, we propose
to adopt the following methodology for the evaluation of architectures,
protocols and algorithms in edge computing systems, described
in~\rsec{simulation}.
On the one hand, we use mininet\footnote{http://mininet.org/} for network
emulation, which we have customized to suit edge computing environments.
Inside mininet, we run actual applications: the overhead of having many
instances on a single \ac{OS} is limited since mininet uses process-based
virtualization, based on Linux's \textit{namespaces}.
The same has been done already in~\cite{Mayer2017}, where the authors
describe their wrapper for an easier configuration of edge/fog
computing topologies, called EmuFog.
On the other hand, we provide a simulator of the computation elements,
which allows large-scale environments to be investigated under limited
availability of hardware resources to execute experiments.
The experiment runs in real time, hence it is possible to mix real
and simulated computation elements seamlessly.
The proposed framework has been used to validate the performance
of the distributed dispatching of serverless functions in~\rsec{eval}.

In~\rtab{sim-comp} we summarize the strengths and weaknesses of the
performance evaluation approaches found in the literature, plus our
framework, in terms of the following key aspects:
\begin{mylist}
  \item accuracy in modeling computation processes;
  \item accuracy in modeling network dynamics;
  \item scalability to large-scale scenarios;
  \item possibility to integrate real applications and components;
  \item capability to obtain consistent and repeatable results.
\end{mylist}
As can be seen, our proposal is an excellent trade-off between
model accuracy and scalability/usability.

This paper is an extended version of~\cite{Cicconetti2019}, including
the following new major contributions: measurement of the average
computational cost of the dispatching algorithm (\rsec{overall-dispatch}),
complete description of the performance evaluation framework
(\rsec{soa:simulation}, \rsec{simulation}); two new batches of
simulation experiments (\rsec{eval:limitations} and \rsec{eval:2kr}).
}

\section{Serverless edge computing}\label{sec:contribution}

In this section we describe the solution envisaged for the execution of stateless tasks, which is most suitable to applications like \ac{AR} or real-time picture/video manipulation, with high computational demands but without a complex state or heavy storage usage (\rsec{contrib:arch}).
Then we delve into the design of the distributed algorithm for dispatching lambda functions to edge servers (\rsec{contrib:dispatch}).

\subsection{Architecture}\label{sec:contrib:arch}

The proposed architecture is illustrated in \rfig{architecture2}.
We consider a generic \ac{MBWA} where \acp{UT} connect to base stations, which are then interconnected though a core network of backhaul network devices.
We assume that devices with computational capabilities, called \textit{computers}, are co-located with the base stations, though not necessarily all of them, and with some of the core network devices.
Such computers, in general, will have heterogeneous capabilities and may be equipped with hardware that is most suitable to execute a specific type of lambda functions, e.g., \ac{GPU} for \ac{AR} and video transcoding~\cite{Albanese2017}.
The base stations, which are the entry point to the network services for clients, act as \textit{dispatchers}.
In practice these base stations could be \ac{LTE} e-NBs or \ac{WLAN} access points or a mix of them.

\myfigeps[scale=0.7]{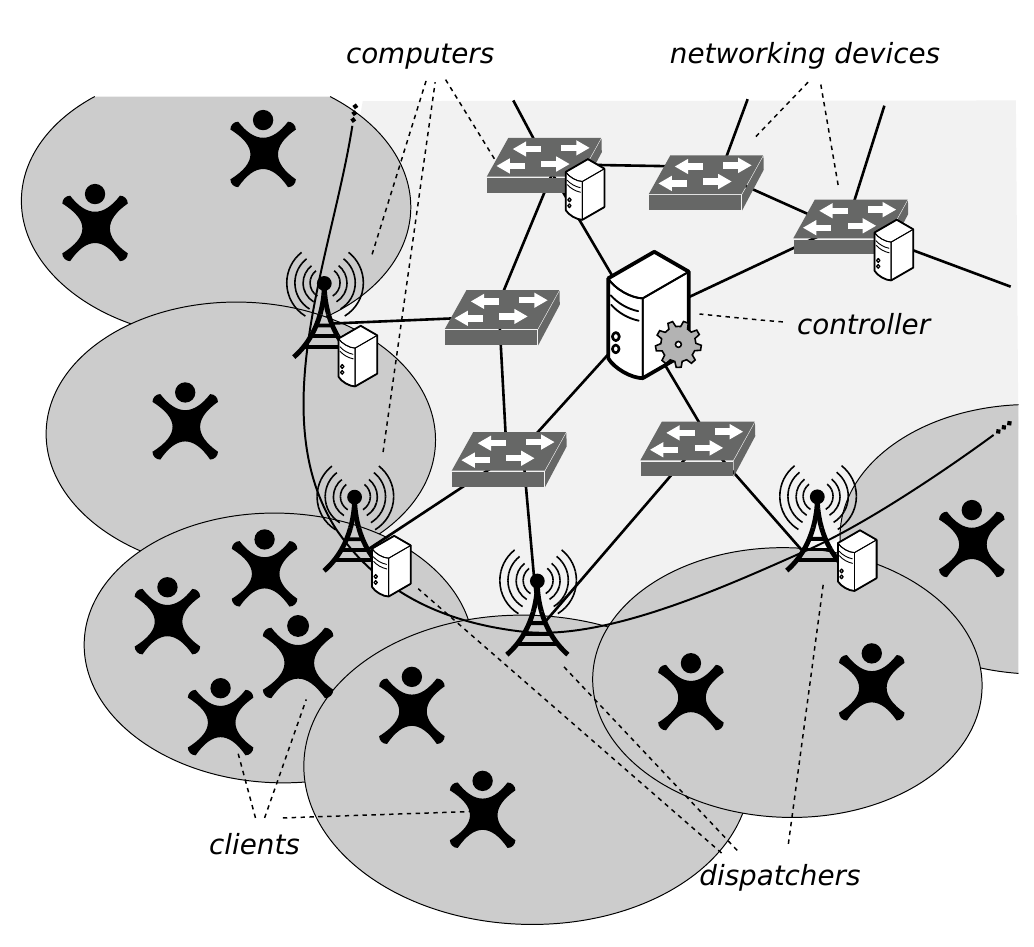}{Proposed distributed system architecture.}

We split the main system functions into two categories: offline and online.
\textit{Offline functions} are performed independently of lambda transactions and are expected to happen on a long time scale (minutes and above).
\textit{Online functions} are those associated to every lambda transaction, thus may happen at very short time scales (seconds and below).
Since latency is a primary concern for pervasive real-time applications, we push as many functions as possible into the offline category: \ac{AA}, set-up of the \ac{VM}/containers on the computers, configuration of the dispatchers.
Such offline functions are enabled by a logically centralized entity, called \textit{controller}, which is represented as a server in the core network in \rfig{architecture2}.
The controller learns about the existence of capabilities of new computers and dispatchers, respectively, and runs a periodic optimization to modify the lambda functions offered by the computers.
In the literature, the latter is referred to as ``service placement'' and some studies have already addressed this topic, e.g.~\cite{Ascigil2017}, far from exhausting it, especially if heterogeneous hardware is considered.
We do not address this issue here, and assume in the rest of the paper that in between consecutive re-organizations the set of \ac{VM}/containers, hence lambda functions, in every computer is stable, thus lambdas can \textit{immediately} be put into execution upon arrival from dispatchers, provided that there is sufficient hardware and software available: e.g., \ac{CPU} and memory, pre-allocated workers and \ac{OS}-related resources.


In \rfig{lambda_sequence} we show the sequence diagram of an online function: the request of the activation of a lambda function from a client, also including the function input, its forwarding to the appropriate computer, and the final communication of the result to the issuing client.
In the next section we discuss the matter of selecting the best computer for the execution of a lambda function, called the \textit{distributed lambda dispatching problem}.

\myfigeps[scale=0.5]{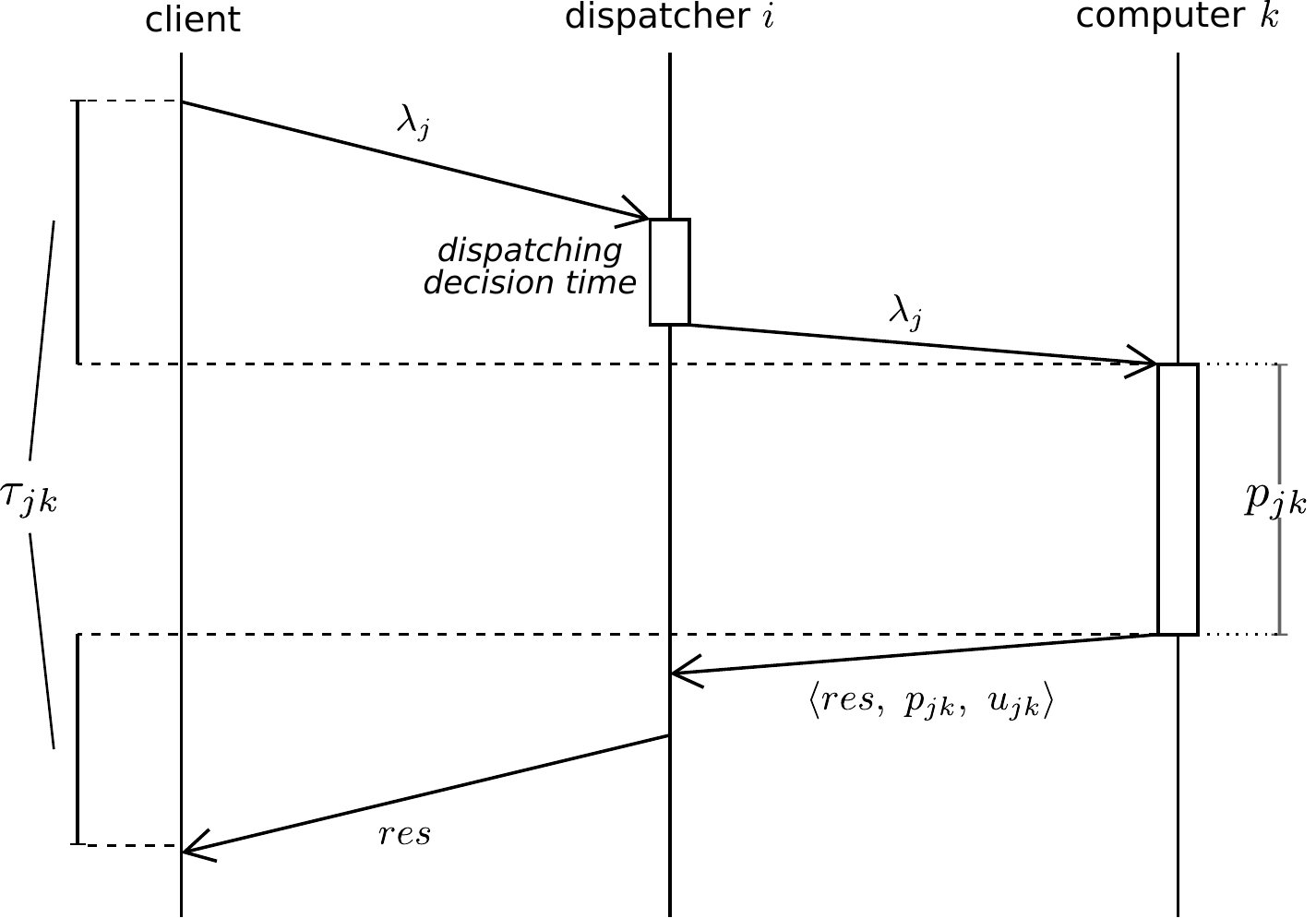}{Lambda request/response sequence.}

\subsection{Distributed lambda dispatching}\label{sec:contrib:dispatch}

We now present the algorithm for selecting the destination of a given lambda function $j \in \mathcal{L}$ ($\|\mathcal{L}\| = L$) at time $t$, provided that there is a set of $\mathcal{C}$ computers that can serve it\footnote{As a recap, this means that the controller, or any other orchestration function in the network, has configured all the \ac{VM}/container/run-time environments necessary for the execution of that lambda function on all computers in $\mathcal{C}$ and that the dispatcher has been informed of such function placement before job $j$ arrives. }, where $\|\mathcal{C}\| > 1$.
We call $\delta_{jk}(t)$ the delay of job $j$ if dispatched to computer $k$ at time $t$.
We can split $\delta_{jk}(t)$ into the following components: $\tau_{jk}(t)$, which is the time required for the transmission of the input from the client to the computer and for receiving the response on the way back, also including all queuing delays in intermediate transmission hops; and $p_{jk}(t)$, which is the time required from processing the lambda function $j$ on computer $k$, which depends on its computational capabilities and other concurrent tasks sharing the resources with $j$ until its completion.
The components of the overall job delay are illustrated in \rfig{lambda_sequence}.
Ideally, the dispatching algorithm should select $\bar{k}$ such that:

\begin{equation}
\bar{k} = \arg\min_k\{\delta_{jk}\} = \arg\min_k\{\tau_{jk} + p_{jk}\},
\end{equation} \\
where we have dropped the time $t$ reference to simplify notation.
This policy is well known in the literature under the name of \ac{SRPT} and is widely employed in multi-server schedulers because of its simplicity.
In addition to having a bounded competitive ratio, it has been also shown to be more resilient than other sophisticated algorithms when the processing time is not certain~\cite{Mailach2017}, which is precisely our case because both $\tau_{jk}$ and $p_{jk}$ cannot be known in advance.
Therefore, we define $\hat{\delta}_{jk}$ as the \textit{estimated} delay of job $j$ if dispatched to computer $k$, and similarly for $\hat{\tau}_{jk}$ and $\hat{p}_{jk}$.
Below we address the research challenge of estimating $\hat{\tau}_{jk}$ and $\hat{p}_{jk}$ in a manner that is
\begin{enumerate*}[label=\roman*)]
  \item \textit{effective}, to emulate as closely as possible the behavior of an ideal \ac{SRPT} scheduler,
  \item \textit{simple}, because the dispatcher has limited resources compared to, e.g., cloud servers in a data center,
  \item \textit{fast}, since we are targeting low-delay applications, therefore we cannot afford to linger too long on the decision of where to direct the lambda function, and
  \item \textit{subject to uncertainty}, for the dispatcher uses only local information, which is bound to become outdated quite fast in a highly dynamic pervasive environment.
\end{enumerate*}

\subsubsection{Communication latency estimation}

As far as $\hat{\tau}_{jk}$ is concerned, we propose a simple, yet effective, mechanism: when a computer is assigned a lambda function, it piggybacks the processing time into the response containing the result.
This allows the dispatcher to sample the communication latency with every computer by simply keeping track of the overall time required for the job execution.
Note that the dispatcher and computer do not need be synchronized since both time intervals are relative.
With some simplifications, we can consider the communication latency as composed of two major components: a fixed offset, which depends only on the network topology and communication technologies used, and a variable quantity that is proportional to the amount of data transmitted.
If we further assume that the lambda function output is either negligible compared to the input or proportional to it, then the dispatcher can collect for every computer a moving window of communication latency samples, obtained from the execution of \textit{any} lambda function, and perform a simple linear regression to derive $\hat{\tau}_{jk}$ once job $j$ arrives, hence its input size is known.
More sophisticated approaches can be used without affecting the core of our contribution.

\subsubsection{Processing time estimation}

The estimation of the processing time $\hat{p}_{jk}$ is more challenging.
In general, predicting the processing time of a non-trivial algorithm executing on a shared general-purpose computer is extremely difficult, because the result depends on a huge number of factors and contingent conditions.
It is beyond the scope of this paper to investigate the issue in full details, as done for instance in~\cite{Pham2017}, where the authors propose a \ac{ML}-based cloud task execution prediction framework.
Furthermore, accurate prediction requires application- and scenario-specific details to achieve best accuracy:
\added{%
for instance in~\cite{Choochotkaew2018} tasklets (see~\rsec{soa:dispatch})
are matched to computation resources based on a learning-based
approach, which however requires the source code of the applications
to be passed through a static profiler for feature extraction.
}
On the other hand, we propose the following practical scheme that
can be used in general cases, i.e., without \textit{a priori}
knowledge of the algorithms (workflow, code, etc.) and the internal
details of computers (\ac{OS}, scheduling policy, etc.).

Firstly, we assume that every computer is able to piggyback on the responses the current system load.
This assumption is rather weak since every modern \ac{OS} is able to provide effortlessly such information to its applications\footnote{Though it may require careful consideration if \ac{VM} or containers are involved since the ``system'' load may not correspond to the achievable load because of virtualization/isolation mechanisms; this is a mere implementation detail, though.}

Secondly, we observe that in practice it is reasonable to expect an increasing relation between the processing time of a given lambda function with given input size and the load in the recent past: if a computer has been heavily loaded in the last few seconds, then it is likely that it will be still so in the near future, thus extending the execution time of any new job assigned.
Also, for a given computer, the processing time will generally increase as the input size increases, all other conditions (e.g., the load) being the same.
Therefore, we propose that every dispatcher keeps track of the past processing times occurred, together with the lambda input size and the load reported by the computer.
This provides all dispatchers with the following 2D mapping for any given lambda and computer:

\begin{equation}
  \langle \mathrm{size, load} \rangle \rightarrow \mathrm{processing~time}
\end{equation}
that can be used to extrapolate $\hat{p}_{jk}$.

In this work we take a simplistic approach and assume that $p_{jk}$ is a linear function of both the lambda input size and the computer load.
Under this assumption, estimation of the processing time can be done by finding the plane that best fits the population of samples collected.
Since this fitting must be updated at every new lambda, we further propose to reduce the computational complexity by quantizing the lambda input size into a set of discrete values, then finding a 1D linear fit as a function of the load values only.
This process is visualized in \rfig{ptime_plane}, which shows the population of processing time values as a function of the lambda input size and load reported collected by a dispatcher during one of the experiments described in \rsec{eval:small} below.
The details of the experiment are irrelevant at this stage, but we anticipate that real computational offloading, i.e., face detection on static images, is being performed.
As can be seen, there are four possible image sizes, yielding an equal number of lambda input sizes, and the processing time increases as either the input size or the load reported increases.
In the 3D plot we show both a linear regression of the 2D plane and four 1D linear regressions, one per lambda input size: the 1D linear regressions are very close to the 2D plane, but they can be achieved at a fraction of the time complexity, which confirms the above working assumption.

\myfigeps[width=3.5in,trim={2cm 1cm 1cm 1cm},clip]{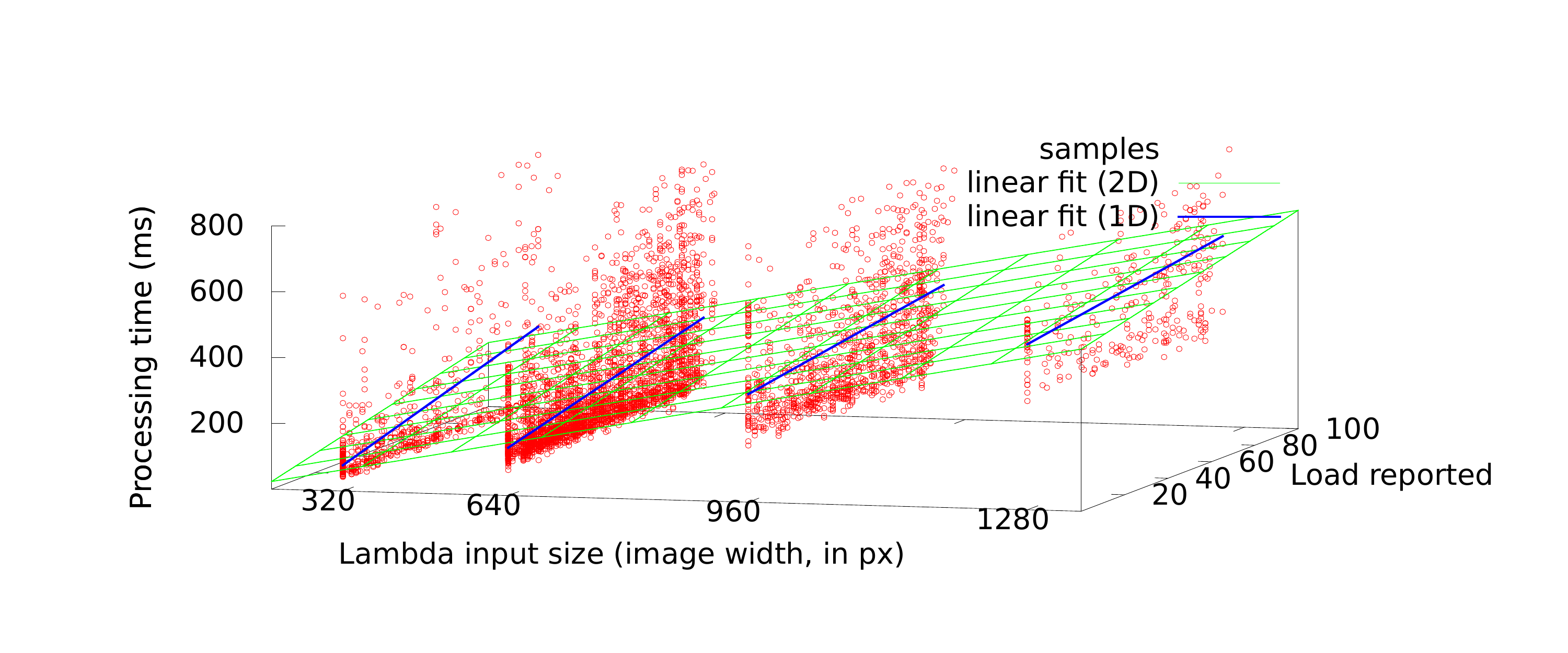}{Example of 2D and 1D linear fitting of processing data values in the dispatcher, for each computer $k$ and lambda function $j$.}

If the assumptions in this section do not apply to a specific scenario, for instance the execution time does not correlate to the input size, then another processing time estimation algorithm (e.g., one more sophisticated or that has white-box knowledge of the applications or computers) can be plugged in seamlessly when implementing the edge dispatching, without affecting the overall framework proposed.

\subsubsection{Overall dispatching algorithm}%
\label{sec:overall-dispatch}

To summarize, every time a lambda function $j$ of size $S_j$ is correctly executed by computer $k$, which reports load $u_k$ and processing time $p_{jk}$, the dispatcher performs the following house-keeping operations:

\begin{enumerate}
  \item measure the communication latency $\tau_{jk}$ as the difference between the overall lambda execution time, which is a local information, and $p_{jk}$;
  \item add $\{ S_j \rightarrow \tau_{jk} \}$ to a moving window of $W_\tau$ samples and find the intercept/slope values $\alpha^\tau_{k}$/$\beta^\tau_{k}$ that best fit them;
  \item quantize $S_j$ as $S'$ to the closest value among the $W_S$ possible ones;
  \item add $\{ u_k \rightarrow p_{jk} \}$ to a moving window of $W_p$ samples and find the intercept/slope values $\alpha^p_{S'jk}$/$\beta^p_{S'jk}$ that best fit them.
\end{enumerate}

\added{%
By design, measurements are always collected as the result of the
dispatching algorithm selecting a computer as the destination of a
lambda function execution request.
This sort of \textit{passive polling} guarantees high scalability
as the number of computers and dispatchers grows, but it has a side
effect possibly leading to sub-optimal selection: once a computer
becomes affected by a bad reputation due to possibly temporary
overload conditions, either due to concurrent tasks or network
traffic, such a status may never be cleared because the dispatcher
is unlikely to select it as the best choice.
To prevent such a starvation, we associate a lifetime to the
measurements: after a timeout $T_\tau$ ($T_u$) the values collected
regarding the communication latency (processing time) estimation
process are discarded, and the computer's state is restored afresh
as if it had just entered the edge computing domain.
In a production environment where an optimization process adapts
the computational/networking resources over time, the lifetime-based
approach could also be substituted by an event triggered on the
dispatchers by the orchestration layer.
%
}

The final dispatching algorithm for an incoming lambda function of
type $j$, whose input is $S_j$, quantized as $S'$, consists of
finding the destination computer $\bar{k}$ s.t.:

\begin{equation}
\label{eq:final-dispatch}
\bar{k} = \arg\min_k\left\{
  \left( \alpha^\tau_{k} + \beta^\tau_{k} S_j \right) +
  \left( \alpha^p_{S'jk} + \beta^p_{S'jk} u_k \right)
\right\}
\end{equation}

To achieve high scalability with non-specialized hardware, it is
important that the dispatching algorithm remains as simple and fast
as possible as the number of computers and lambda functions grow.
The \textit{worst-case computational complexity}, in both space and
time, of the main algorithm components is reported in \rtab{complexity},
where the \textit{house-keeping} rows refer to operations that are
carried out upon receiving a successful response from a computer.
We briefly recall the notation used in the table: $L$ is the number
of possible lambda functions, $C$ is the number of computers in the
edge network, and $W_\tau$, $W_S$, and $W_p$ are the internal
parameters representing the number of communication latency samples
kept per computer, the number of quantized input sizes, and the
number of processing times kept per lambda per computer, respectively.

\begin{table}[htbp]%
\caption{Computational complexity analysis.}%
\begin{center}%
{\small%
\begin{tabular}{|r|r|r|}
\hline
\textbf{Algorithm} & \textbf{Space} & \textbf{Time} \\
\hline
Communication latency house-keeping & $\mathcal{O}(W_\tau C)$ & $\mathcal{O}(W_\tau)$ \\
\hline
Processing time house-keeping & $\mathcal{O}(L W_p W_S C)$ & $\mathcal{O}(W_p)$ \\
\hline
Dispatching  & -- & $\mathcal{O}(C)$ \\
\hline
\end{tabular}
}%
\label{tab:complexity}%
\end{center}%
\end{table}%

\added{%
On the other hand, to quantify the \textit{average computational
complexity} of the algorithm we have carried out the following
testbed experiments.
We have run a single instance of a dispatcher, whose internal
structure and implementation details are illustrated in~\rsec{sim:dispatcher},
on a Raspberry Pi~3 Model~B (RPi3), which is ``representative of a
broad family of smart devices and appliances''~\cite{Morabito2017a},
and on an Intel application server with Xeon CPU E5--2640 v4 at 2.40GHz.
Even though the application can exploit parallelism on \ac{SMP}
architectures by spawning multiple threads, for the purpose of the
experiment we have restricted the execution on a single core\footnote{This
can be done in Linux using the so-called \textit{affinity} mechanism.}.
In the experiment we have installed a given number of possible destinations
of a special lambda, for which the dispatcher replies immediately to the
client without actually reaching a computer.
However, all other phases of the dispatching algorithm, including the
computation of~\req{final-dispatch} and the update of all the
internal data structures and timers, are performed exactly as with
regular lambda functions.
We have run experiments with 1 and 4 threads spawned in the dispatcher,
and executed a matching number of clients repeatedly asking for the
special lambda function to be executed.

\myfigeps{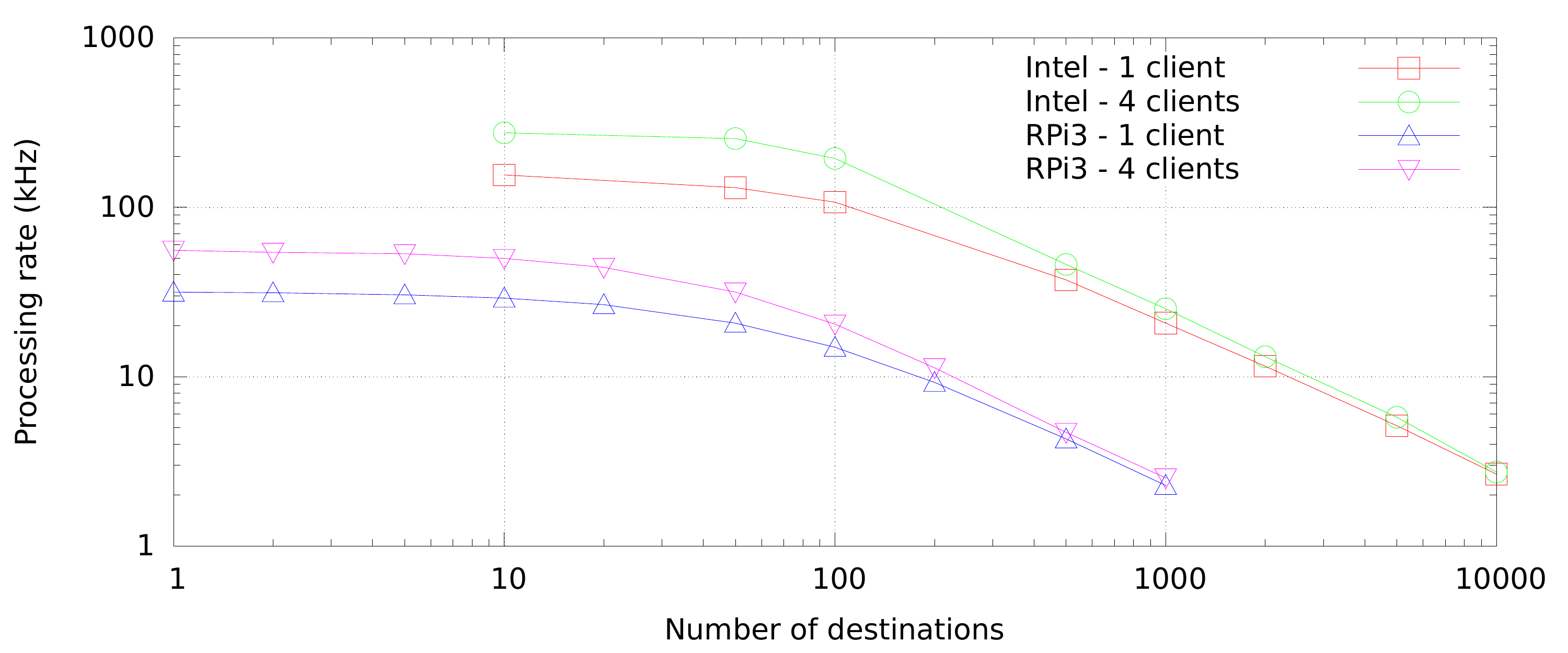}{Dispatcher processing rates.}

In~\rfig{rate} we report the processing rate achieved in the different
combinations, which gives an upper bound of the performance of the
dispatcher (per single core) and a quantitative measure of the
computational cost of the proposed algorithm as the number of
destinations ($C$) increases.
For each combination of parameters we have executed 10 repetitions,
but we do not report error bars in the plot because the variance
was negligible.
As can be seen, in all cases the processing rate decreases linearly
in the log-log plot, which means that the average computational
complexity increase w.r.t $C$ is sub-linear, unlike the worst-case
computational complexity, which is in fact often considered too
pessimistic.
This provides for a smooth scalability of the dispatcher as the
edge computing size, i.e., the number of computers, increases.
In absolute terms, the Intel server-grade \ac{CPU} obviously performs
much better than RPi3's ARM \ac{CPU}, and achieves a processing
rate significantly greater than 1000~Hz, which means less than 1~ms
overhead per function call, with up to 10,000 possible destinations.
However, also the RPi3 incurs a reasonable small overhead in a range
that is certainly relevant for many practical deployments.

In the supplementary material we also include the \ac{CPU} utilization
measured during the experiments.
}


\section{Simulation framework}\label{sec:simulation}

\added{%
As illustrated extensively in~\rsec{soa:simulation}, to the best
of our knowledge, the existing solutions to evaluate the edge
computing performance have some limitations when realistic modeling
of both \textit{connectivity} and \textit{computation} is required.
In this section we propose a novel framework that overcome these
limitations.
The framework has been developed originally for the performance
analysis of the serverless edge computing architecture described
in~\rsec{contribution}, but it can be used ``as is'' for the
evaluation of generic algorithms and protocols in a wide range of
scenarios.
}

\added{%
Our solution builds on network emulation and lightweight
virtualization, combined with simulation of the computation processes.
Therefore, all system components are instanced as Linux applications
running inside \texttt{lxc} containers, which is a form of process-based
virtualization, incurring a negligible overhead compared to running
the same application on the host \ac{OS}.
This allows the experimenter to take into account in a realistic
manner all the effects due to the communication protocols in use,
as well as including in the analysis several phenomena that are
very difficult to capture using simulators, e.g., inter-process
communication overhead and caching.
Furthermore, network emulation can be realized efficiently
using mininet, which is known to scale well to large
topologies with thousands of nodes on a single server\footnote{If
this is not sufficient, its extension MaxiNet
(\url{https://maxinet.github.io/}) allows the distribution of mininet
``islands'' over multiple servers, thus enabling theoretically to
scale up experiments to any arbitrary size, provided that sufficient
computational and network capabilities are available.}.

The use of real components in experiments, however, does not mean
that necessarily \textit{all} applications must be real ones.
In fact, we argue that in many cases the use of real edge applications
is not needed.
This is the case of our distributed dispatching system for serverless
edge computing, described in~\rsec{contribution}, which focuses on
the forwarding of the functions, but it is agnostic of the actual
applications being run by the clients.
Since edge applications are very likely to be a choke point for the
execution of large scale experiments, because offloading
is especially appealing for computationally-intensive tasks that
cannot be executed efficiently on devices, by simulating computation
we can retain all the advantages of a real testbed with a fraction
of the hardware required.
This, in turn, opens the door to easy automatization of the execution
of the experiments, simplified collection/analysis of results, and
straightforward repetition of experiments.
}

%
\added{%
In the remainder of this section we first describe the experiment
workflow (\rsec{sim:workflow}) and the computation element simulation
model (\rsec{sim:sim-computer}), which are both general and
application-agnostic.
Afterwards we focus on the serverless-specific components of our
framework: we illustrate the image manipulation computer
(\rsec{sim:real-computer}), used in~\rsec{eval} directly, i.e.,
instances are executed as part of the experiments, and indirectly,
i.e., to calibrate the configuration parameters of the computation
element simulation model, and finally the serverless edge computing
dispatcher (\rsec{sim:dispatcher}).
}

\added{%
\subsection{Experiment workflow}\label{sec:sim:workflow}

\myfigeps{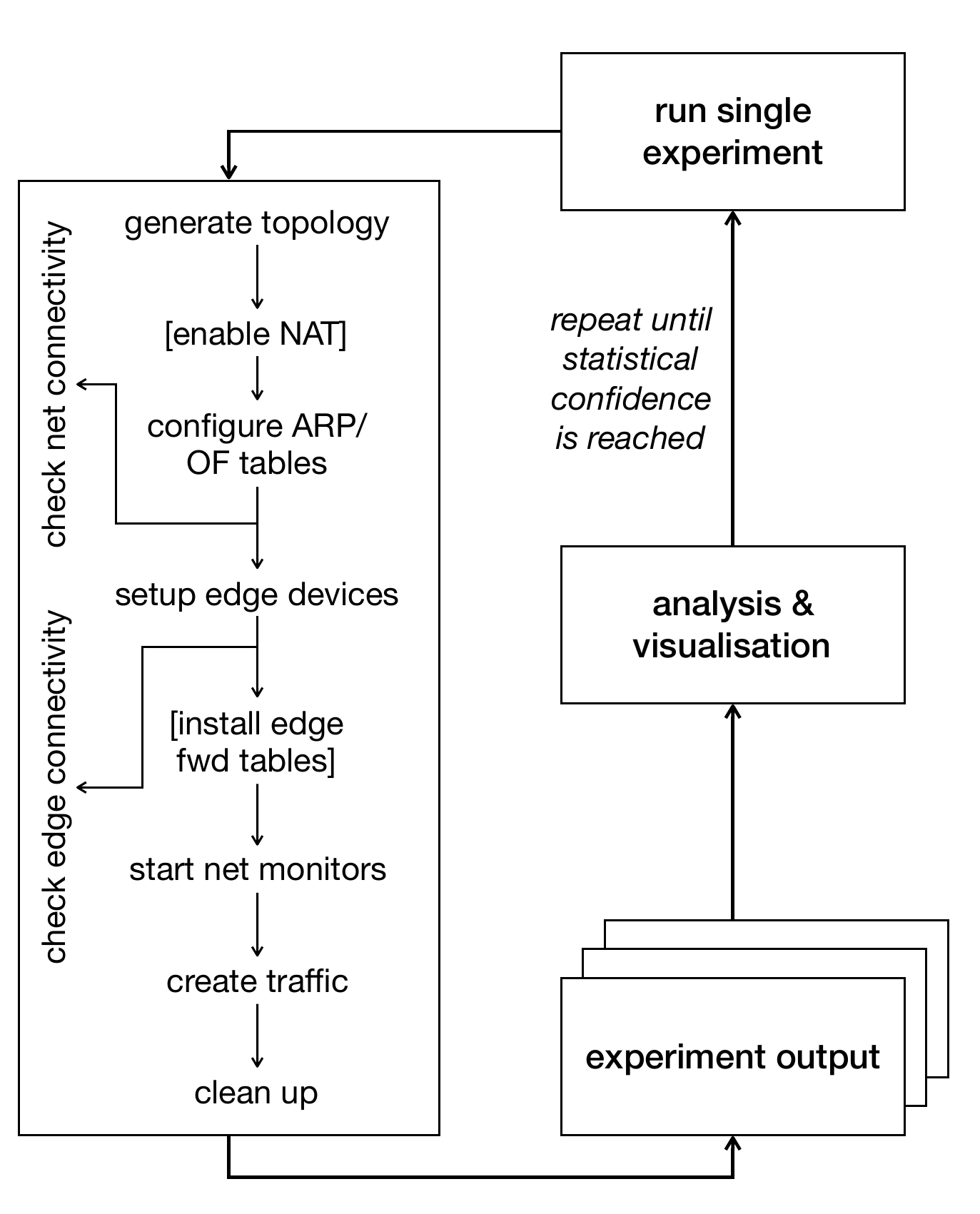}{Workflow of a generic experiment.}

In~\rfig{flow} we illustrate the generic experiment workflow: the
execution of a single experiment is done by launching a Python
script, which includes a number of generic modules developed as
part of our framework, interacting with mininet via a Python \ac{API}.
The modules implement the logic to realize all the steps reported
in the left hand side box in~\rfig{flow}, including the topology
generation according to models suitable for edge computing environments
and the configuration of the \ac{OF} and \ac{ARP} tables in switches
to ensure network-layer connectivity.
Every intermediate step may be customized by the experimenter by
overriding the relevant methods of the main experiment class to
suit their specific needs.
Common-use metrics, such as per-switch throughput, are collected by
default and made available at the end of the experiment, together with
custom metrics.
The analysis and visualization of the experiment output is instead
beyond the scope of the framework, since it ultimately depends on
the actual environment.

Our framework is general enough that it can be used to simulate a
multitude of different edge computing environments, also including
interactions with real world applications living outside the
virtualized mininet environment through an appropriate \ac{NAT}
configuration.
We have used the framework to validate and assess the performance
of our distributed architecture for serverless edge computing
with real and simulated computation elements, in a vast range of
network topologies, using different methodologies (transient
analysis, steady-state with replications, Monte Carlo).
}

\subsection{Computer simulation model}\label{sec:sim:sim-computer}

We now present the model of the \textit{simulated computer}, which
is an application that performs arbitrary functions
without really implementing any algorithm, but rather simulates
internally the execution of the currently scheduled lambda functions
to mimic the behavior of an \ac{SMP} multi-container edge server.
Tasks are served according to a \ac{FCFS} non-preemptive
policy.
\added{%
Every task is associated to requirements in terms of both computation
(number of operations) and memory (bytes).
In the simulated computer implemented we have used a linear model
where the number of operations (or memory) required is the sum of a constant
offset and a value that is proportional to the input of the lambda
function request, in bytes.
In the supplementary material we show that this model is able to
capture very well the response time of the image manipulation
computer described in~\rsec{sim:real-computer}.
Different offset/slope values can be used for different lambda
functions in different computers: this allows to simulate computers
that have heterogeneous characteristics, such as equipped with
\acp{GPU}/\ac{VPU} better suited to graphics/\ac{ML} jobs
than general-purpose \acp{CPU}.
The number of operations is used to determine the simulated
processing time, which also depends on the number of cores and
the \ac{CPU} speed.
Multiple tasks can share the available cores, provided that
the container on which they are running have sufficient
\textit{workers} available, otherwise the tasks are put into
a waiting list.
On the other hand, the memory required is used to block tasks whose
requirement would exceed the residual memory available, which is
the difference between the total amount installed (in simulation)
and the sum of the memory requirements of all active tasks.
If a task is blocked because of insufficient memory, no other task
is put into execution until it is eventually made active, to avoid
starvation.
Different models to compute the computation/memory requirements,
as well as different scheduling policies, can be easily implemented,
should they be needed to better model a given platform or application
under evaluation.

%

\myfigfulleps%
{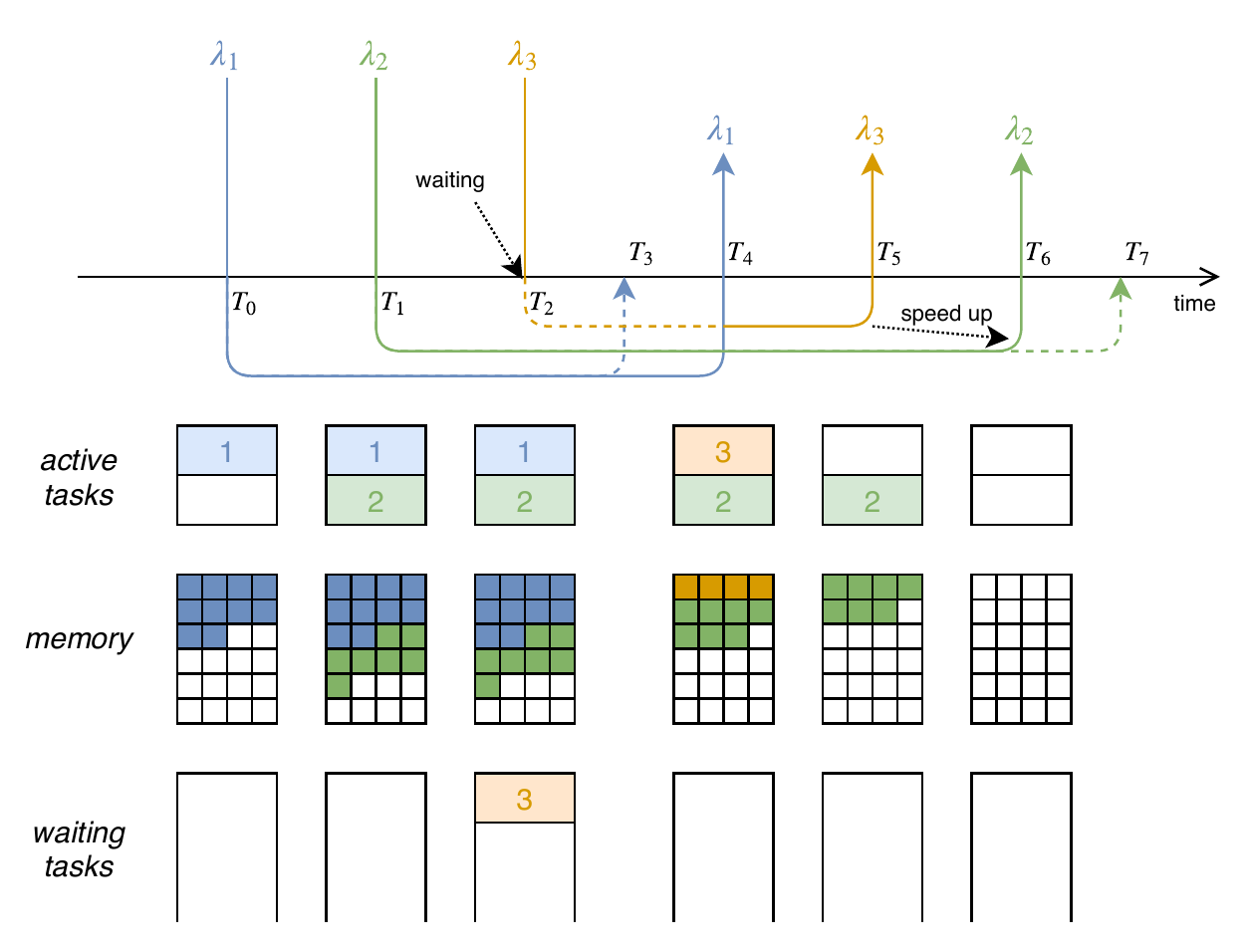}%
{Example of processing simulation.}

To better explain the behavior of the simulated computer we now
illustrate two example simulations, both with three incoming tasks.
For simplicity we assume there is a single simulated core.
All the lambda functions are served by the same container, i.e., they
are of the same type, with two workers.
In the first example, in~\rfig{computer-seq}, memory is not a
limitation and it is not considered (it is the subject of the
second example).
At $T_0$ the first task $\lambda_1$ enters the system; based on its
input size and the offset/slope configured for this container in this
computer, the completion time $T_3$ is assigned: if no other tasks
arrived, that would be the time when $\lambda_1$ returns.
However, at $T_1$ a second task $\lambda_2$ arrives: in addition
to assigning a completion time to it ($T_7$), the simulated computer
also modifies the completion time of $\lambda_1$ to $T_4$ because from now
on it has to share \ac{CPU} cycles with $\lambda_2$.
When a third task $\lambda_3$ arrives, it is put into the waiting
list because there are no workers available; this situation changes
only as $\lambda_1$ leaves the system, when $\lambda_3$ is put
immediately into execution: note that the completion time of
$\lambda_2$ which is still active, does not change because the
number of \textit{active} tasks remains the same.
On the other hand, as $\lambda_3$ completes its execution, $\lambda_2$
becomes the only active task, which makes the computer shorten its
completion time from $T_7$ to $T_6$.

\myfigfulleps%
{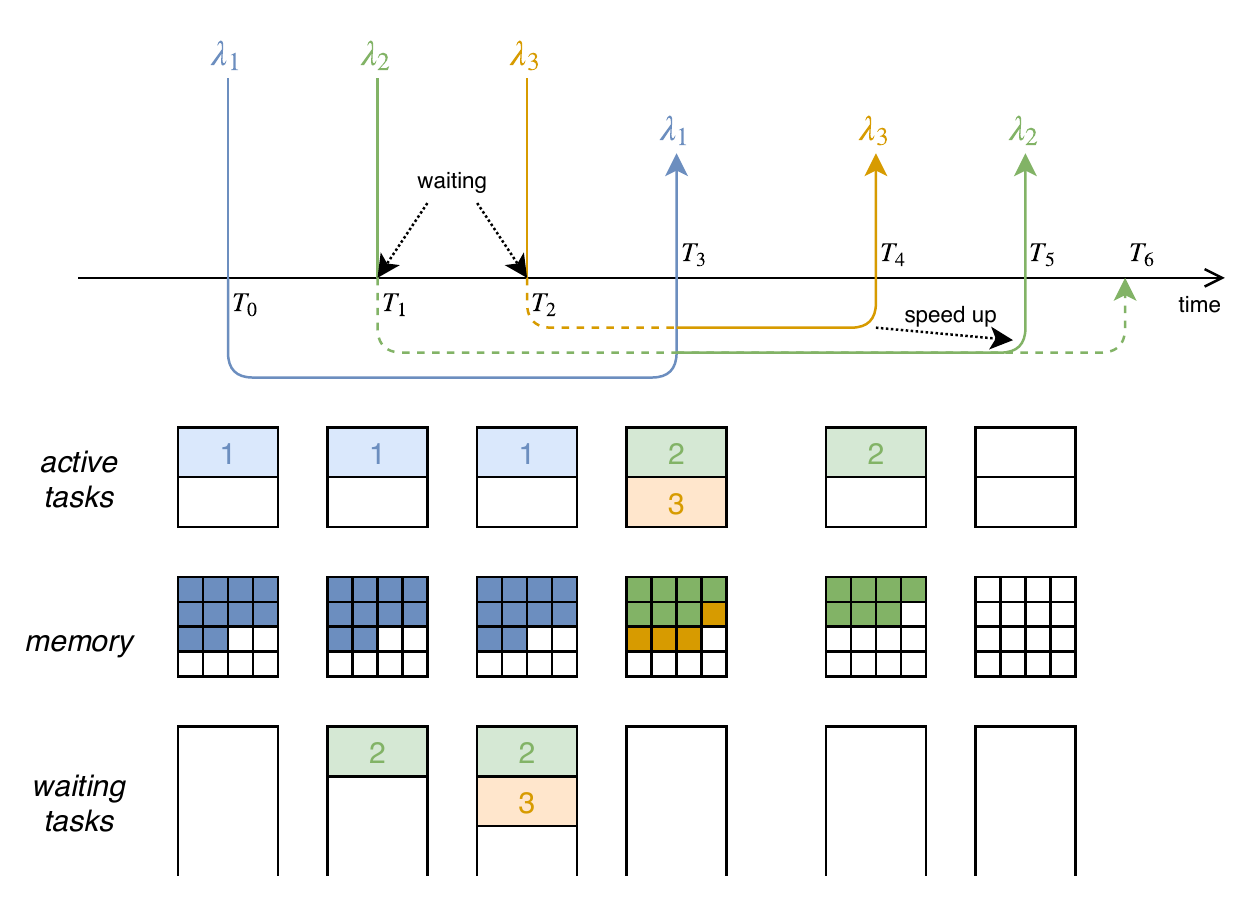}%
{Example of processing simulation, with memory-bound tasks.}

In the second example, in~\rfig{computer-seq-mem}, the tasks have
exactly the same characteristics and arriving times, but we now
have a smaller memory.
Because of this tighter constraint, as $\lambda_2$ arrives at $T_1$,
it cannot be made active because memory is insufficient.
Therefore it is put into the waiting list, while $\lambda_1$ enjoys
full \ac{CPU} power, hence no completion time advance like in the
first example.
When $\lambda_3$ arrives at $T_2$, its memory requirements \textit{would}
fit into the residual capacity, but it is put into a waiting list
together with $\lambda_2$ because we enforce a \ac{FCFS} policy.
As already mentioned, this also automatically avoids starvation of
tasks with large memory requirements.
As soon as $\lambda_1$ completes at $T_3$, both waiting tasks become
active, and the simulation then proceeds as in the first example.

\myfigeps{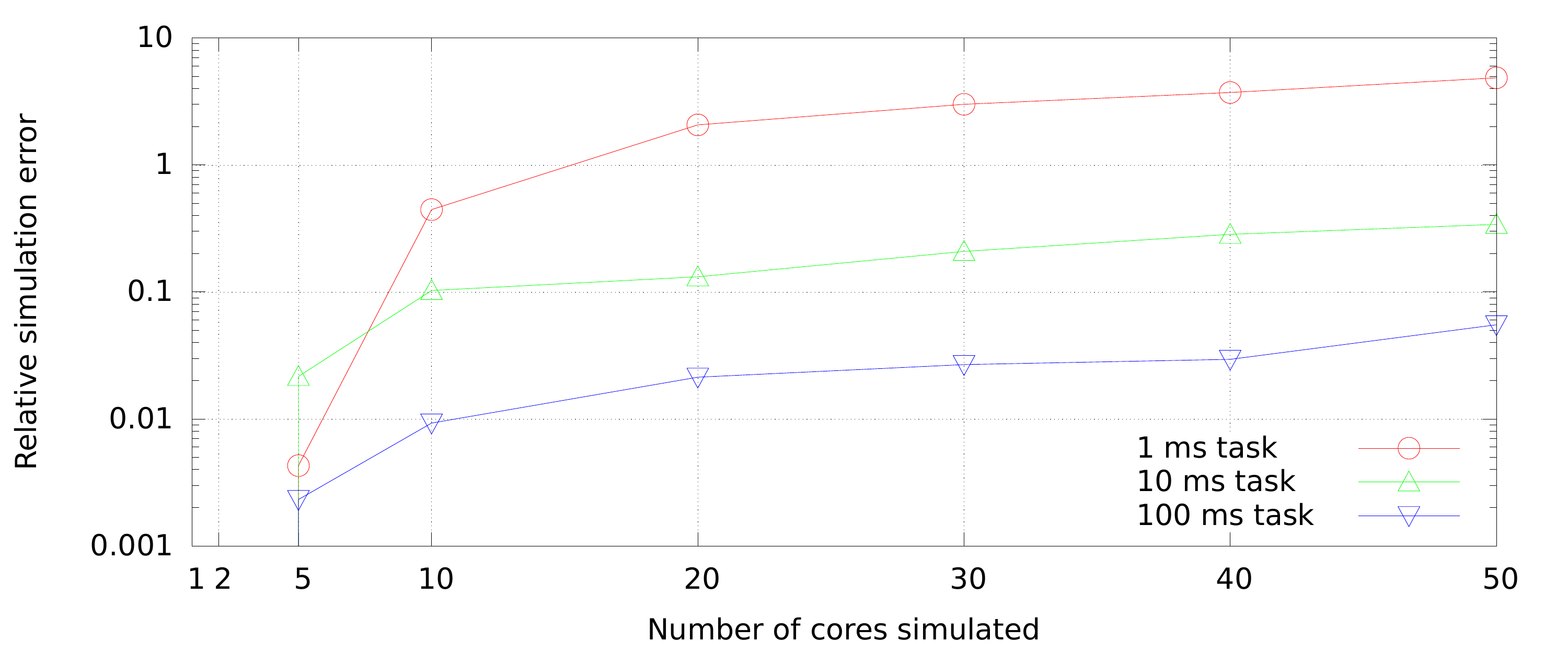}{%
Relative execution error of the computer simulator as the number
of simulated cores increases.}

Quite clearly, since the simulation must happen in real-time,
there are limitations to the processing rates that can be achieved
by the simulated computer.
We now show in a quantitative manner, with the results from an
experiment, that the range of operation of the simulated computer
are very reasonable.
We run an instance of a simulated computer on a single core
of an Intel Xeon CPU E5--2640 v4 at 2.40GHz (same machine used
for the dispatcher results in~\rsec{overall-dispatch} above
and for the experiments in~\rsec{eval} below).
We are interested into measuring the error introduced by
the simulated computer as a function of: the number of cores
simulated and the processing rate.
To this aim, we set up a number of clients equal to the number of
simulated cores, continuously requesting the execution of a lambda
function configured with offset/slope such that its processing time
is 1~ms, 10~ms, and 100~ms, respectively.
Memory limits do not constrain execution of applications.
In these conditions, we expect that the response time will be exactly
the same as the nominal processing time, because every task has
a dedicated simulated container/core for its execution: every deviation
is due to the \textit{physical} \ac{CPU} of the host not being
able to keep the pace with the simulated events.
In~\rfig{threads-error} we plot the \textit{relative execution
error}, defined as the ratio between the actual processing time and
the expected processing time minus 1.
Since the actual processing time is always greater than the expected
value, because a \ac{CPU} may only lag in time, the relative execution
error is always positive: for instance a value of 0 means that there
is perfect match between simulated and actual processing times, whereas
a value of 0.1 means that the computer is introducing an error in
the order of 10\% of the processing time.
In the supplementary material we also report the physical \ac{CPU}
utilization during the whole experiment.
As can be seen in the plot, a single physical core is able to
simulate with small error a server equipped with many cores, up to
50 in our experiments, provided that the processing time is 10~ms
or higher.
For instance, this means simulating $50 \times 20 = 1000$ containers
on the server we have used, which has 20 physical cores.
On the other hand, if simulation with fine-grained granularity of a multi-core
machine serving such shortly-lived tasks is required, it will be
necessary to allocate sufficient physical \ac{CPU} cores to the
computers to obtain accurate results at high processing rates.
}

As already introduced, this simulated computer is very important
for performance evaluation purposes:
\begin{enumerate*}[label=(\roman*)]
  \item it allows to scale experiments up to large networks without
  requiring prohibitive computational capabilities,
  \item it enables the execution of sensitivity analysis studies,
  having a totally known and controllable response, and
  \item finally it allows the implementation of the comparison
  algorithm proposed in~\cite{Tan2017} because it can predict the
  completion time of jobs (if no other arrives meanwhile).
\end{enumerate*}

\subsection{Image manipulation computer}\label{sec:sim:real-computer}

\added{%
In addition to the simulated computer described above, for the
purpose of evaluating our serverless edge computing
architecture with a realistic application of practical interest,
i.e., mobile \ac{AR},
}
we implemented an \textit{image manipulation computer} that actually
performs face or eyes detection using the OpenCV
library\footnote{\url{https://www.opencv.org/}}.
This also shows how our performance evaluation framework can work
with real applications, simulated ones (as described
in~\rsec{sim:sim-computer} below), or any mix of the two.
%
%
However, we are not interested in the details and challenges
associated to the specific detection algorithms, for which we refer
the interested reader to, e.g.,~\cite{Wang2017}.

\added{%
The OpenCV library is able to use concurrently multiple \ac{CPU} cores
to reduce the processing time.
The maximum concurrency level can be set at run-time, and in the
experiments we use this feature to artificially limit the computational
capabilities of computers.

With regard to face detection, the edge client on the device sends
a picture to the dispatcher selecting \texttt{face\_detection} as the
lambda name.
The dispatcher then forwards the request to a suitable computer,
which replies with the rectangles enclosing all the faces found,
if any.
The response time is the time between when the edge client issues the
lambda request and when it receives back the response.
If detection of eyes is also requested, then the edge client, for
every face found in the first step, crops the original image based
on the rectangle coordinates and issues another lambda request of
type \texttt{eyes\_detection}, which is also dispatched as before
to a suitable computer.
In this case the response time is measured by the client between
when the initial face detection is requested and when the last eyes
detection response is received.

\begin{table}[htbp]%
\caption{%
Response times of the lambda functions used in the experiments with
an image manipulation computer for face detection.}%
\begin{center}%
{\small%
\begin{tabular}{|r|r|r|r|}
\hline
  \textbf{Picture size} & \textbf{Size (bytes)} & \textbf{Comp.\ only (ms)} & \textbf{With network (ms)} \\
\hline
   320$\times$240 &  52830 &  43 $\pm$  9 &  60 $\pm$ 10 \\
   640$\times$480 & 175332 & 101 $\pm$ 10 & 218 $\pm$ 26 \\
   960$\times$720 & 353230 & 181 $\pm$ 13 & 446 $\pm$ 47 \\
  1280$\times$960 & 560103 & 301 $\pm$ 13 & 744 $\pm$ 45 \\
\hline
\end{tabular}
}%
\label{tab:opencv-times}%
\end{center}%
\end{table}%

We report in~\rtab{opencv-times} the response times of face detection
with the set of pictures used in the performance evaluation, ranging
from 320$\times$240 to 1280$\times$960, when using up to two \ac{CPU}
cores.
We report the times both when called directly (\textit{Comp.\ only}
column in the table) and when the lambda function is invoked by a
client to a computer via a network link (\textit{With network}
column in the table) emulating a typical \ac{WLAN} access network.
}
The variance of results is rather high due to the \ac{ML} algorithm
used in the OpenCV library for detection.

\added{%
\subsection{Dispatcher prototype}\label{sec:sim:dispatcher}

\myfigfulleps%
{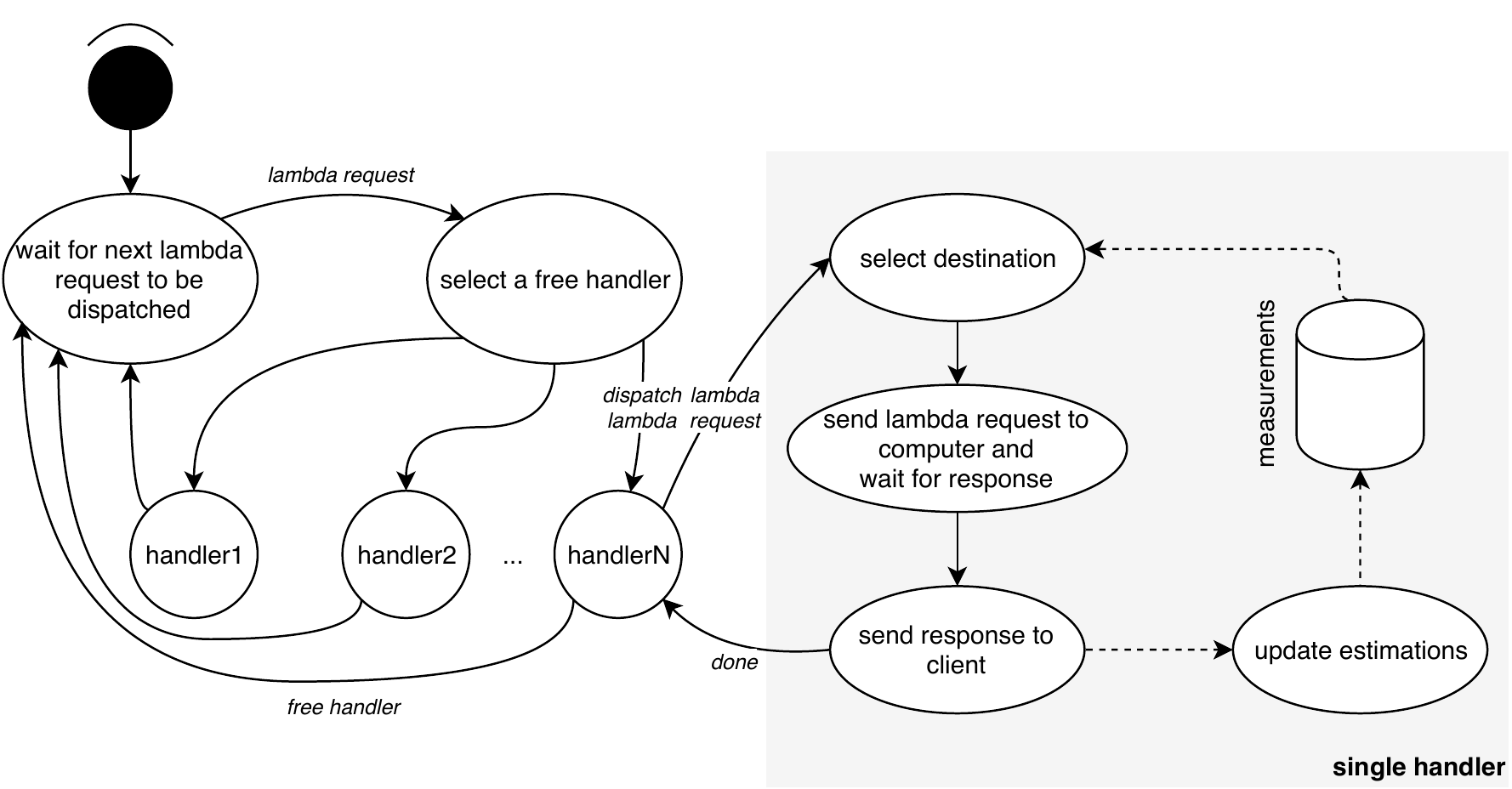}%
{Finite state machine of a dispatcher.}

We conclude this section by describing the online lambda function
dispatcher implemented in our framework, which can be represented
using the \ac{FSM} in~\rfig{router-fsm}.
As illustrated, the dispatcher continuously waits for new
lambda execution requests from clients.
Once a new one arrives, one of the idle handlers takes care of
its entire execution until a response is returned to the client
before it can be considered idle again.
A handler initially selects the destination based
on~\req{final-dispatch}, then forwards the lambda request to the
computer selected and waits from a response, which is eventually
forwarded to the client.
Afterwards, the handler uses a locally measured delay together with
the processing time and load indication piggybacked by the computer
on the response to perform the so-called house-keeping operations
in~\rsec{overall-dispatch}.
Both the latter and destination selection use data structures that
are in common for all the handlers, which may limit in principle
the degree of parallelism that can be achieved as the number of
handlers is increased, especially for lambda functions with a short
execution time.
}

We implemented the dispatcher in C++ and the execution of lambda
functions has been realized by means of REST interface methods using
Google's gRPC\footnote{\url{http://grpc.io/}}, which is a mature
industrial-grade communication protocol using \ac{HTTP}, with
messages serialized with Google's protobuf library%
\footnote{https://developers.google.com/protocol-buffers/}, which
is lightweight and portable.
A \textit{lambda request} contains: the name, that is used to
identify the container or run-time environment to be used; the
input, that is opaque to the edge computing components
\added{%
; and a flag that, if enabled, tells the dispatcher not to actually
forward the lambda but reply immediately with an estimate of the
time it would be required to carry out the function.
This last field could be used by a client associated to multiple
dispatchers to decide where to request execution.
Think for instance of a user with a smart phone connected to a
mobile network and a \ac{WLAN}, both offering edge computing services.
}
%
A \textit{lambda response} contains: a return code specifying what
went wrong, if anything; the output; the \ac{URL} of the computer
actually carrying out the computation; the time required for the
execution of the lambda; the response, opaque to the edge computing
components; a short-term average load of the computer before the
execution of the lambda function.

In the prototype, our applications, written in C++, call directly
the REST methods on the dispatcher to which they are attached, but
integration with any other high-level programming language is
possible since the gRPC library is platform and language independent.
To simplify the discovery phase, we assumed that every computer and
dispatcher registers itself to the controller at a known \ac{URL}.
Finally, to achieve interoperability in a real deployment we envisage
adopting standard \acp{API}, such as those defined by the \ac{ETSI}
\ac{MEC}~\cite{Schiller2018}; such an opportunity is being currently
investigated.

\section{Performance evaluation}\label{sec:eval}

\added{%
In this section we evaluate the distributed lambda dispatching
algorithm with the proposed performance evaluation framework in
three setups: an artificial scenario in \rsec{eval:limitations},
configured in different flavors to show the flexibility of our
simulation framework to model very heterogeneous conditions, as well
as to introduce the main characteristics of the proposed distributed
dispatching algorithms; a small-scale network of edge nodes, equipped
with real face detection capabilities in \rsec{eval:small} (this
set-up is also used for an assessment of the sensitivity of results
to the main internal parameters in \rsec{eval:2kr}); a more realistic
large-scale environment where we have used simulated computers
(\rsec{eval:large}).

The experiments have been run on a Linux Intel Xeon dual socket
workstation, with Intel Hyper Threading disabled, that was not used
by any concurrent demanding application.
During all the experiments the computational processing demands
were never significant enough to suggest that the results might be
affected by a hardware limitation of the host used.
In all the experiments we have used $W_\tau = W_p = 100$  based on
preliminary calibration experiments: we do not report these results
because they would give the reader little insight, however we have
verified \textit{a posteriori}, using the method of $2^k r!$
analysis~\cite{law2007simulation}, that the results are not affected,
in a statistical sense, by setting these parameters to 50\% and
200\% of the value above, see \rsec{eval:2kr} below for full details.
}
The value of $W_S$ depends on the specific experiment and is indicated
in the respective sections below.
%
%
Unless otherwise specified, every experiment has been repeated 10
times: in the plots we report the 95\% confidence interval over all
the replications.

\added{%
For all the scenarios below, additional results are available as
part of the paper supplementary material.


\subsection{Exploring resource limitations}\label{sec:eval:limitations}

\myfigeps[width=2in]{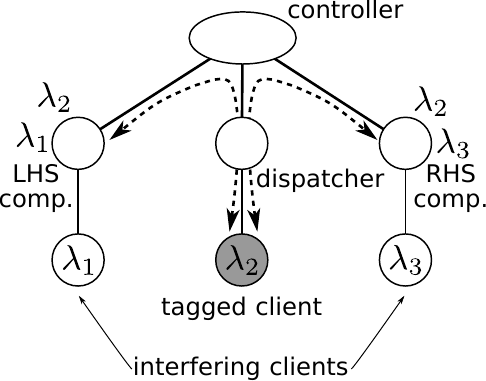}{Network topology used in the resource limitations experiments.}

In this section we show that our proposed framework can adapt to a
wide range of environments: as described in \rsec{simulation} this
is very important for the performance evaluation in edge computing
environments, since the latter span across several markets / vertical
industries, each with very specific and heterogeneous characteristics.
We use the simple network topology illustrated in \rfig{het-exp-topo},
in which the thick edges have 100~Mb/s bandwidth / 1~$\mu$s latency,
and they connect a root node (also hosting the controller) to the
edge nodes, whereas the thin edges have 25~Mb/s bandwidth / 100~$\mu$s
latency, connecting the clients to their respective edge nodes.
The \ac{LHS} computer offers serverless functions $\lambda_1$,$\lambda_2$
and the \ac{RHS} computer offers $\lambda_2$,$\lambda_3$; every
client issues lambda function requests of the same type indicated
in the figure: as a result of these assumptions, the so-called interfering
clients on the left and right of the network may be served only by their
respective computers, while the tagged client in the middle has its
requests served by the \ac{LHS} or \ac{RHS} computer depending on the
decisions of its dispatcher.

We have configured the computers to give the same response times, on average,
as a real image manipulation computer for face detection, reported
in \rtab{opencv-times}.
The tagged client issues lambda request with size corresponding to
a 320$\times$240 picture every 200~ms on average (actual inter-time
between consecutive requests is drawn from a uniform r.v.); on the
other hand, the interfering clients, when present, use all possible
picture sizes in the table with an average inter-time of 1~s.
We simulate four cases:
\begin{mylist}
  \item Baseline: without interfering clients, only to establish
  a reference of the performance metrics;
  \item CPU: we make the two computers unbalanced in terms of
  computation power by reducing to $\frac{1}{3}$ the number of RHS's
  simulated cores;
  \item Memory: we make the two computers unbalanced in terms of
  memory by specifying a limited memory for RHS, while the memory
  of LHS never constrains execution of tasks in this scenario;
  \item Network: we simulate background traffic between the root
  node and the LHS Computer by transmitting bursts of TCP exchanges
  lasting for 3~s every 5~s.
\end{mylist}

In all the four cases we compare the following schemes:
\begin{description}
  \item[Probe] (inspired from \cite{Tan2017}): a centralized dispatcher
  co-located with the controller polls each computer to retrieve
  the execution time required, then selects the one that advertised
  the smallest value; the implementation of Probe with simulated
  computers (see~\rsec{sim:sim-computer}) is straightforward since
  they always know precisely the execution time of any incoming
  task, provided that no other jobs arrive.
  \item[RPI] (Relative Performance Index, inspired from
  \cite{Breitbach2019}): the dispatcher uses a \ac{WRR} scheduler
  to select the destination of incoming lambda function requests,
  with the weight being equal to its zero-load response time,
  normalized in range whose bounds are given by the fastest and
  slowest computer, respectively; the weights are computed in an
  initial profiling phase during which the dispatcher executes every
  lambda function on every computer, while no other task is being
  executed on it.
  \item[Est]: our proposed dispatching algorithm described
  in \rsec{overall-dispatch}.
\end{description}
Clearly, both Probe and RPI cannot be implemented in a real deployment
and they are to be considered only as optimistic performance
reference: on the one hand, Probe requires the execution time of
any task to be known \textit{a priori}, but this information is
generally not available; on the other hand, the initial profiling
of RPI requires that the edge computing system is unavailable
throughout this phase, which may be unfeasible in a live system.

\myfigeps{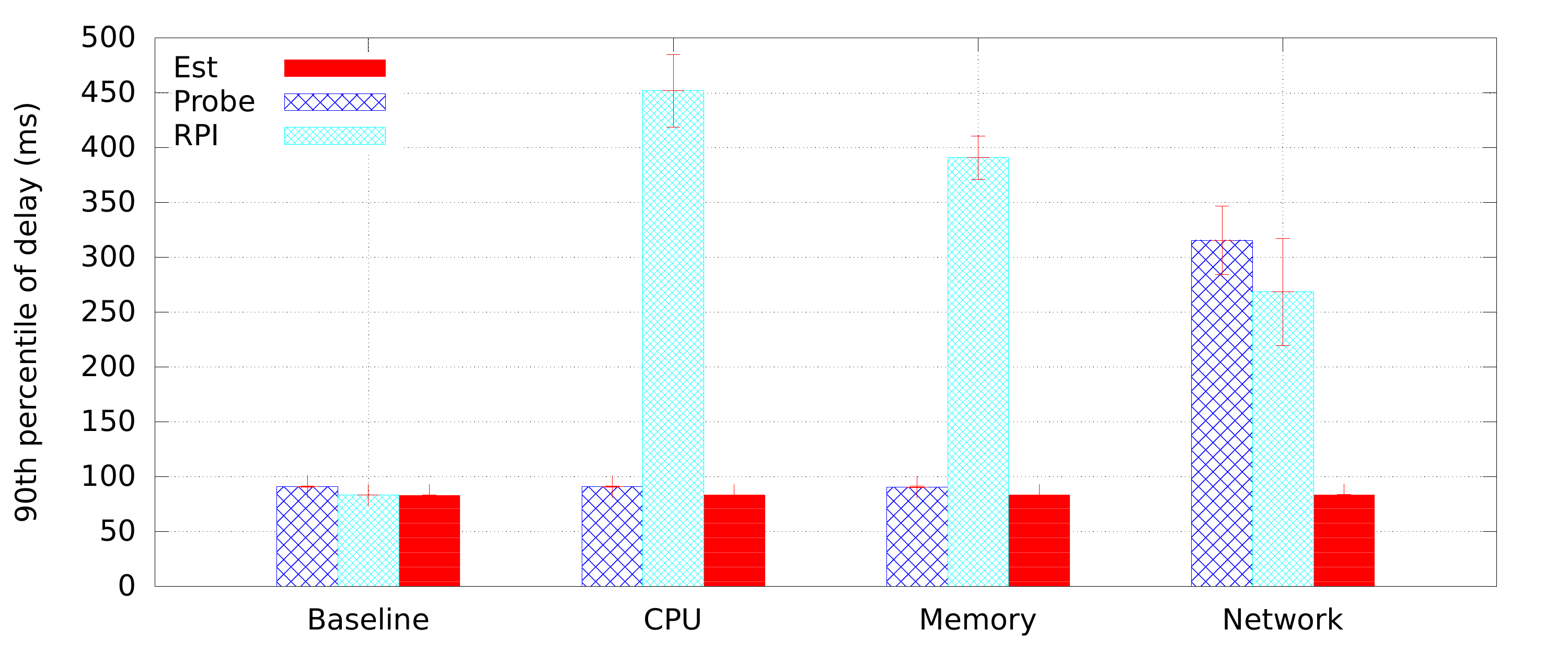}{Resource limitations: 90th percentile of delay.}

In \rfig{het-out-90th} we show the 90th percentile of the delay,
which is defined as the time between when the client issues a lambda
request and when it receives the response from the dispatcher.
In the Baseline case the delay is the same for RPI and Est, but
slightly greater with Probe, due to the additional delay of polling
the computers to check which one would provide the shortest processing
time.
In the CPU and Memory cases, the 90th percentile of delay increases
significantly with RPI because the latter blindly balances between
the two computers, regardless of their current state.
On the other hand, our proposed dispatching algorithm, using only
local estimated information, is as efficient as a centralized
version that uses reliable information from the computers themselves.
In the Network case, Est outperforms both comparison schemes since
it also includes in its algorithm (see~\req{final-dispatch}) the
communication latency.

\myfigeps[width=\textwidth]{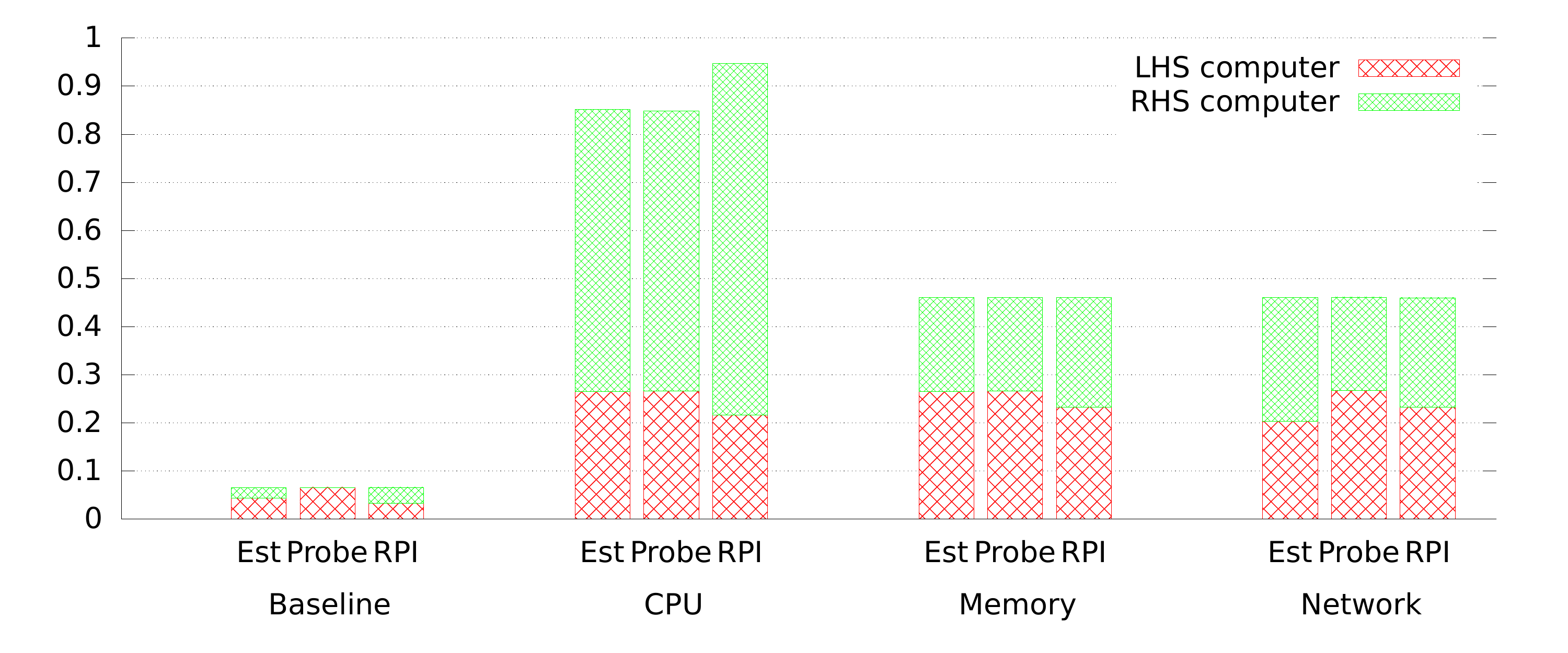}{%
Resource limitation: average load.}

In \rfig{het-util-split-mean} we provide additional insight on the
system dynamics by showing the average load of the computers, defined
as the fraction of time that the simulated cores are actively
processing tasks.
First, we note that with Baseline the Probe scheme uses only one
computer: since they are identical and there are no other clients
than the tagged one competing for resources, polling deterministically
returns the same value on both computers.
Second, in the CPU case, the RHS load is higher because it has less
simulated computational resources.
Finally, in the Memory and Network cases, the LHS computer's load
is respectively the same with both Probe and RPI: this is because
both solutions take into account only the processing time, either
statically identified in a profiling phase or dynamically requested
from the computers, which yields worse performance than Est, which
instead is more flexible in adapting to the different computational
capabilities and serverless/background traffic conditions.
}


\subsection{Small-scale experiments}\label{sec:eval:small}

In this set of experiments we arranged four edge nodes interconnected
in a clique with 100~Mb/s links with 1~$\mu$s latency.
This is representative, for example, of a local edge environment
supporting a group of mobile nodes, deployed by a \ac{MEC} operator
through a set of edge gateways located close to each other.
Each edge node hosts an image manipulation computer that can perform
face detection via the execution of lambda functions as described
in \rsec{simulation}.
The computers are assigned different capabilities: the computer on
node $i$, with $i \in [1..4]$, can use up to $i$ \ac{CPU} cores
among those available in the server hosting the experiments.
All clients connect to the edge nodes via links with 25~Mb/s capacity
with 100~$\mu$s latency.
For the convenience of evaluation we measure the latency of ``tagged''
clients only, one per edge node, that issue, on average, one lambda
request per second with picture size 640x480 pixels according to a
Poisson distribution.
The number of other clients, requesting detection on pictures
randomly drawn from a set of images from 320x240 to 1280x768 pixels
and roaming from one edge node to another selected randomly, increases
from 1 to 4.
In these experiments it is $W_S = 4$.

\myfigeps[width=2in]{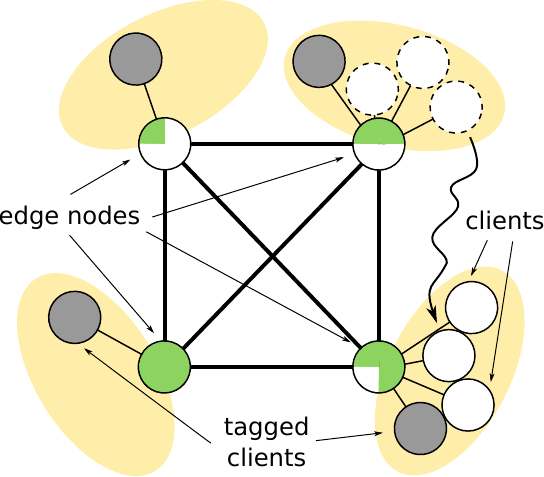}{%
Network topology used in the small-scale testbed experiments.}

We compare the performance obtained with our proposed dispatching
solution, called Est in the following, with two alternative approaches.
First, a \ac{RR} algorithm, taken from our previous
work~\cite{Cicconetti2018}, which classifies the computers based
on lower vs.\ higher response time, then dispatches the incoming
lambda request to the computers that are currently in the lower
category.
\ac{RR} does not distinguish between communication and processing
delays and does not use the load values reported by the computers.
Second, we consider a Legacy approach, where the clients simply
request the execution to the closest computer, in number of hops.

In \rfig{opencv_delay_90th} we show the 90th percentile of the delay
of tagged clients.
As can be seen, at low network loads \ac{RR} performs worse that
the others, because it strives to use evenly the available computers,
which however have different capabilities.
This behavior pays off at high loads, where, on the other hand,
Legacy is penalized because it cannot cope well with ``hot spots''
of clients.
In all cases the Est curve lies always below the others: our proposed
approach can adapt well to mixed environments.
Recall that this is achieved in a fully distributed manner and
without providing Est with any \textit{a priori} knowledge on the
topology and capabilities of computers.

\myfigeps{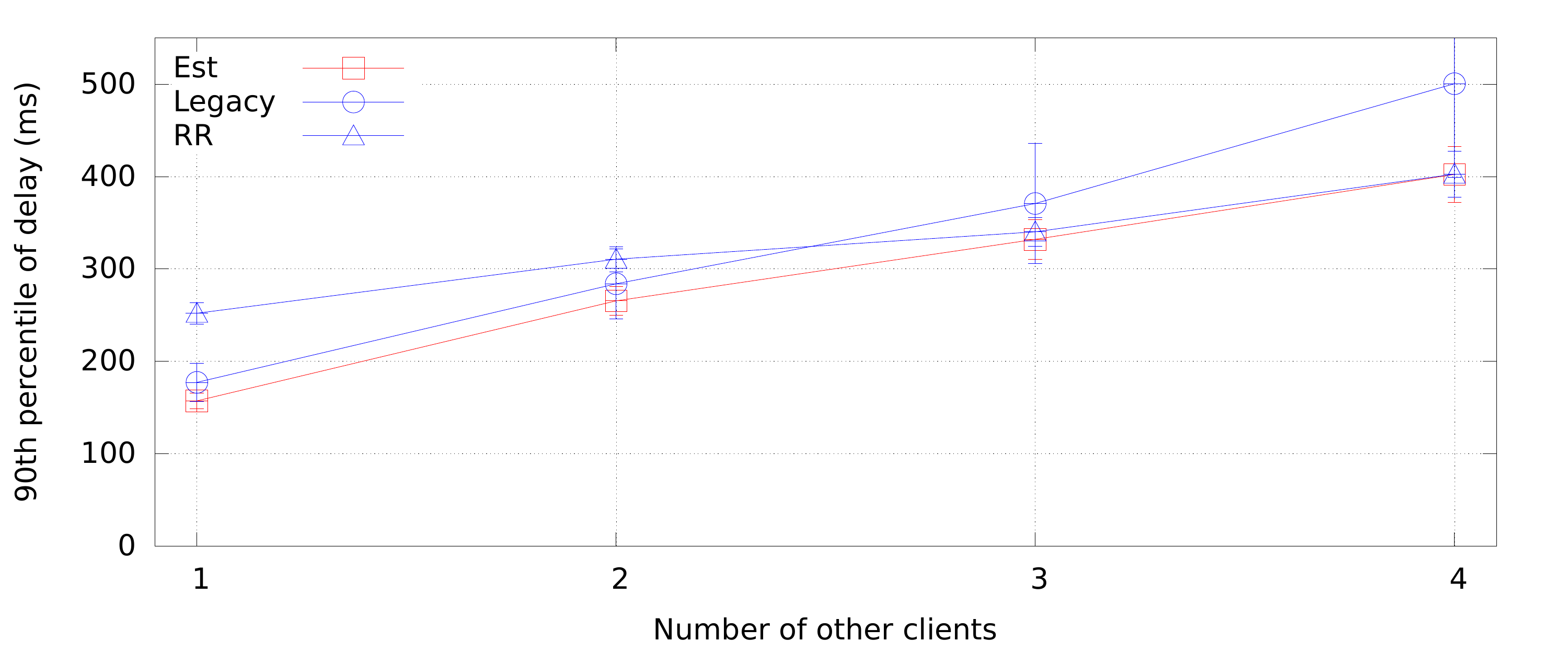}{Small-scale experiment: 90th percentile of delay.}

The reason is explained with the use of \rfig{opencv_util_2}, which
shows the average load of the computers at peak load.
As can be seen, Legacy has an almost flat utilization, which is
inefficient since the load is evenly distributed across edge nodes
but the capabilities are not.
\added{%
Note that a condition of unbalanced capabilities is far from being
artificial: rather, in a real environment it is very likely that
hardware and software on edge nodes, unlike their cloud
counterparts, will be highly heterogeneous due to incremental
deployment and fragmented ownership.
}
On the other hand, both Est and \ac{RR} distribute the load
proportional to where there are more resources.
For instance, even though computer 1 has lowest capabilities, it
has non-negligible utilization with Est, which is thus able to
harvest resources even from less powerful computers as necessary.

\myfigeps{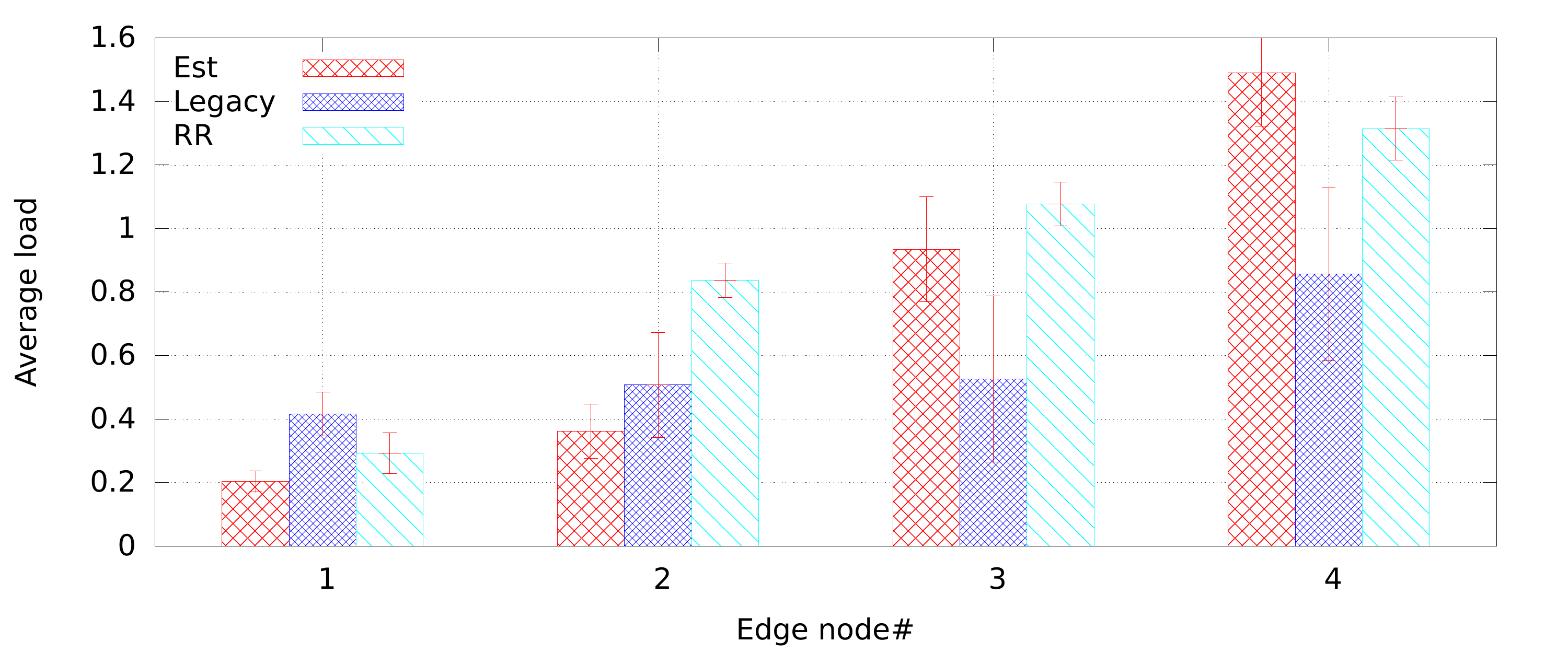}{%
Small-scale experiment: Distribution of load across edge computers,
with four other clients.}


\added{%
\subsection{Sensitivity analysis}\label{sec:eval:2kr}

In this section we study the sensitivity of the proposed dispatching
algorithm to variations of its internal parameters.
We carry out this analysis using a $2^k r!$ analysis, which is an
established methodology to identify the dominant factors in a
statistically sound manner and, usually, plan experiments accordingly.
Very briefly, for each of the $k$ factors \textit{potentially}
affecting the performance, we identify two limit cases, identified
by a $+$ and $-$ sign, respectively.
Then, we run $2^k$ experiments in all possible combinations.
We also repeat $r$ times every experiment, without modifying the
parameters.
For every metric of interest, the $2^k r!$ method allows to derive
the relative importance of every factor or combinations of factors,
also quantifying the contributions in the same units as the performance
index under study, within a given confidence interval.
One basic assumption for this method to provide meaningful results
is that the factors identified give additive contributions to the
metric of interest (though the analysis may still be carried out
with non-additive contributions with some modifications).
The interested reader may find additional information on this
methodology in~\cite{law2007simulation}.

\begin{table}[htbp]%
\caption{Sensitivity $2^k r!$ analysis: factors.}%
\begin{center}%
{\small%
\begin{tabular}{|l|c|c|c|}
\hline
Parameter & & $-$ & $+$ \\
\hline
Number of other clients & A & 1 & 4 \\
$T_\tau$ & B & 1 & 10 \\
$T_u$ & C & 1 & 10 \\
$W_\tau$ & D & 50 & 200 \\
$W_u$ & E & 50 & 200 \\
Detect eyes & F & no & yes \\
\hline
\end{tabular}
}%
\label{tab:2kr_params}%
\end{center}%
\end{table}%

The sensitivity analysis is carried out in the same topology
as \rsec{eval:small}.
The $k = 6$ factors identified are reported in~\rtab{2kr_params}
and they include internal parameters ($W_\tau$, $W_u$, $T_\tau$,
$T_u$, see \rsec{overall-dispatch}) as well as the following
environmental parameters: the number of other clients, which is
basically a measure of the computational/network load; and whether
the serverless clients request only face detection or also the
detection of eyes within every face found in the pictures.
In the latter case all the computers offer two lambdas, one for
face and the other for eyes detection, and the delay is the response
time defined in \rsec{sim:real-computer}.
We performed the analysis in terms of the following metrics: delay,
network throughput, load of the computers, and communication
latency/processing time estimation error, defined as the absolute
value of the difference between the \textit{a priori} estimation and
the \textit{a posteriori} measurement done by the dispatcher for every
lambda function.
We run 10 independent replications for every combination of parameter,
i.e. $r = 10$, yielding a total of $2^6 \cdot 10 = 640$ experiments
executed.

\begin{table}[htbp]%
\caption{%
Sensitivity $2^k r!$ analysis: 90th percentile of delay (ms).}%
\begin{center}%
{\small%
\begin{tabular}{|ccc|}
\hline
  \textbf{Effects} & \textbf{Contribution} & \textbf{Conf.\ int.} \\
\hline
  mean response (q0) & $342$ & $(340, 343)$ \\
\hline
  effect of A (qA) & $+75.6$ & $(73.7, 77.5)$ \\
  effect of F (qF) & $+43.4$ & $(41.5, 45.3)$ \\
  other effects & \multicolumn{2}{c|}{\textit{negligible}} \\
\hline
  unexplained variations & \multicolumn{2}{c|}{9\%} \\
\hline
\end{tabular}
}%
\label{tab:out_90th}%
\end{center}%
\end{table}%

In~\rtab{out_90th} we show the analysis in terms of the 90th
percentile of delay, produced by the open source tool
\textit{factorial2kr}
used\footnote{\url{https://github.com/ccicconetti/factorial2kr}}.
%
%
%
Confidence intervals are computed with 95\% level.
From the table we see that the 90th percentile of delay is affected
merely by environmental parameters A (number of other clients) and
F (detection of face only vs.\ face+eyes).
The sign of the qA and qF values also gives us the direction, which
is as expected: the latency increases as the number of other clients
increases from 1 to 4 (the average increase is 75.6~ms) and if we
also include eyes detection (the average increase is 43.4~ms).
%
%

%
\begin{table}[htbp]%
\caption{%
Sensitivity $2^k r!$ analysis: median of processing time estimation
error (ms).}%
\begin{center}%
{\small%
\begin{tabular}{|ccc|}
\hline
  \textbf{Effects} & \textbf{Contribution} & \textbf{Conf.\ int.} \\
\hline
  mean response (q0) & $22.2$ & $(21.8, 22.6)$ \\
\hline
  effect of A (qA) & $+13.4$ & $(13.0, 13.8)$ \\
  effect of F (qF) & $-19.7$ & $(-20.1, -19.3)$ \\
  joint effect of A\&F (qAF) & $-12.6$ & $(-12.9, -12.2)$ \\
  other effects & \multicolumn{2}{c|}{\textit{negligible}} \\
\hline
  unexplained variations & \multicolumn{2}{c|}{4\%} \\
\hline
\end{tabular}
}%
\label{tab:ptime_50th}%
\end{center}%
\end{table}%

In~\rtab{ptime_50th} we report the analysis when considering the
processing time estimation error $|\hat{p}-p|$.
As for the 90th percentile of delay, only the environmental factors
have non-negligible impact on the performance.
However, there are two differences.
First, the sign of qA and qF is opposite, which means that in this
case the processing time estimation errors \textit{increases} as the load
increases, but it \textit{decreases} if the clients also perform
eyes detection.
This is an effect due to the shorter duration of eyes detection
compared to face detection, because when requesting the latter the
clients only include as input the portion of picture corresponding
to a single face.
Second, a combination qAF appeared in the table: this means that
21\% of the processing time error is due to the \textit{joint}
effect of the number of other clients and application of choice,
which can be explained by the fact that having more serverless
traffic also gives more measurements to the dispatcher, which can then
be more precise in their estimations, especially with faster detection
algorithms.

For all the metrics considered, we have verified that the results
are valid in a statistical sense by visually inspecting the residual
errors, whose residuals vs.\ predicted values and Q-Q normal plots
are reported in the supplementary material.
}


\subsection{Large-scale topology}\label{sec:eval:large}

In this section we report the results obtained with a large-scale
realistic network topology, where lambda functions are executed on
simulated computers.
The target application is \ac{AR} on mobile devices in a dense urban
environment.
We use as reference real-world traces available as open datasets
and described in \cite{Barlacchi2015}, which include recordings of
human activity in the city of Milan (Italy) for one month.
The city landscape is divided into a square grid of 10,000 cells.
We assume that each cell contains a base station serving users
divided into three sectors.
Base stations are grouped into sets of three elements, forming
so-called pods, that are then connected to a common core.
The resulting network topology is a fat-tree, commonly found in
data-centers and operator core networks~\cite{Al-Fares2008}, where
the network links between the root node and its children have 1~Gb/s
with 10~$\mu$s latency, whereas capacity is halved in the tier
below.
The links connecting the clients to their respective base station
have a 25~Mb/s capacity with 1~ms, which is a slightly optimistic
estimate with respect to actual findings in current 4G networks
\cite{Bui2017}.
The mapping of the city grid into an emulated network for the
experiments is illustrated in \rfig{milano_topo}.
\added{%
In particular, the network view (right side) shows the connectivity
graph, where the thickness of the edge is a qualitative indicator
of the link speed: at the terminals level, one node represents all
the user terminals in a sector; at the base stations level, one
node represents a single (sectorized) base station, which acts as
both a dispatcher and a computer, with two cores entirely dedicated
to processing lambda functions; at the pods level, one node represents
the collection of networking equipment creating a backhaul connection
between the access and core networks; finally, the entire core
network is collapsed into the root node of the tree.
}

\myfigfulleps{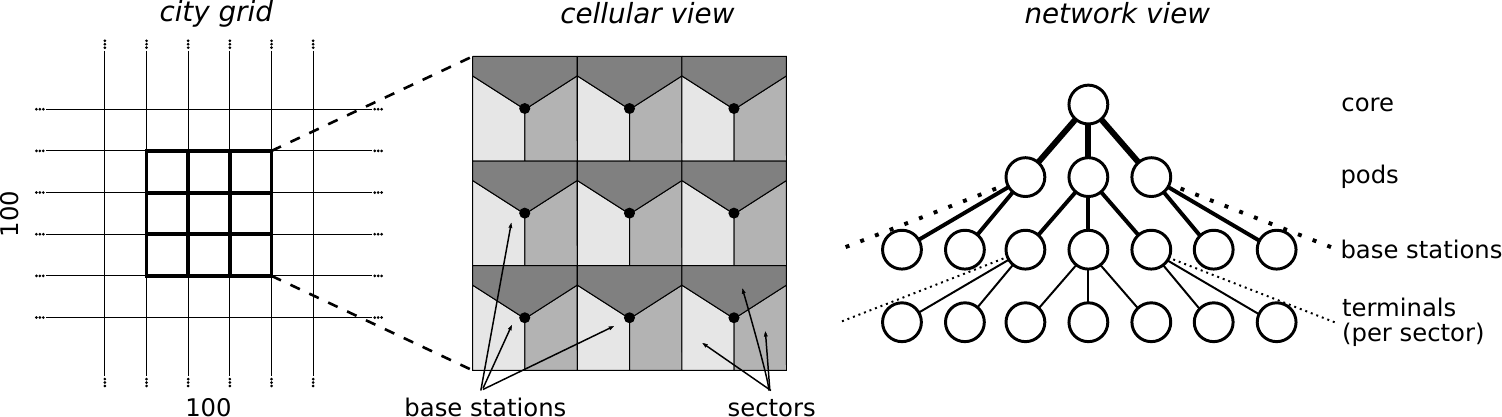}{Mapping of the city grid from~\cite{Barlacchi2015} into the network topology used for the experiments.}

For the experiments we used a Monte Carlo approach, which is widely employed for system-level performance evaluation of \ac{MBWA} algorithms and protocols and is carried out as follows.
Inspired from the evaluation in \cite{Ceselli2018}, from a random day in the dataset in~\cite{Barlacchi2015} we extracted Internet activity with a 10-minute granularity, that is used to determine the cell load at every given time of day.
Then, we perform a number of independent snapshots (or drops) of the system\footnote{In the other experiments we have measured and plotted confidence intervals through the execution of multiple independent replications of the very same scenario. With a Monte Carlo approach the notion of ``independent replication'' is blurred because every drop represents a possible state of the system at a given time, and any two drops may capture very different conditions (e.g., night-time vs.\ peak hours). Therefore, rather than taking averages and reporting some measure of the variance, as customary with the method of independent replications, we report the full results obtained in all the drops using distributions.}.
For each snapshot we select a random location of a group of 3x3 cells and a random time of day.
Then, we drop users with random arrival times with a rate that is proportional to each cell activity at the given time.
Each user establishes a session of \ac{AR} with a duration randomly extracted between 30~s and 60~s, consisting of a stream of lambda function requests directed to the dispatcher co-located with the serving base station.
During a session, consecutive lambda functions are issued every 33~ms, which corresponds to a frame rate of 30~fps.
The size of lambda requests/responses is such to have bandwidth demands ranging from 3~Mb/s (lambda size 5000~bytes) to 10~Mb/s (lambda size 15000~bytes), which according to the authors in \cite{Braud2017} is a reasonable compromise for good quality \ac{AR} under realistic network constraints.
The response times of the lambda function with some image sizes are reported in \rtab{emul-times}.
We consider 75~ms as the maximum tolerable round-trip delay for the execution of a lambda function~\cite{Braud2017}.
In the dispatchers we used a value of $W_S$ such that lambda sizes are quantized every 1000~bytes.

\begin{table}[htbp]%
\caption{%
Response times of the lambda functions used in~\rsec{eval:large}
with a simulated computed emulating an \ac{AR} application.}%
\begin{center}%
{\small%
\begin{tabular}{|r|r|r|}
\hline
  \textbf{Size (bytes)} & \textbf{Comp.\ only (ms)} & \textbf{With network (ms)} \\
\hline
  5000 bytes & 9.0 $\pm$ 0.2 & 12.1 $\pm$ 0.4 \\
  10000 bytes & 17.5 $\pm$ 0.2 & 24.1 $\pm$ 0.3 \\
  15000 bytes & 25.9 $\pm$ 0.2 & 35.9 $\pm$ 0.5 \\
\hline
\end{tabular}
}%
\label{tab:emul-times}%
\end{center}%
\end{table}%

We compare our proposed solution, called Dist Est (= distributed with processing estimation) below, to the following alternatives.
First, a centralized approach where the dispatcher is located in the root node of the topology tree, which mimics the behavior of a typical serverless solution, such as OpenWhisk. 
Second, a distributed version of the Probe algorithm, hence called Dist Probe, described in \rsec{eval:limitations}.
Third, like in \rsec{eval:small}, Legacy, where the clients request the execution of lambdas to their serving base station.

In \rfig{milano_delay_90th} we report the 90th percentile of the delay experienced by the fraction of users in the x-axis.
For instance, with Dist Est we have a value of 40~ms at 0.7 users: this means that 70\% of the users, at every random location and time of day, experienced a 90th percentile of delay that is smaller than 40~ms.
Therefore, we can compare these values to the 75~ms target and find the fraction of dissatisfied users as the complement to 1 of the x-axis projection of the point where each curve meets 75~ms.
In this respect, Legacy achieves poorest performance: this is because a static allocation leads to under-utilization of the computational resources, which is consistent with the main finding in~\cite{Malandrino2016}.
This is mainly due to the fact that with Legacy we create ``hot spots'' of requests at each base station whenever a high concentration of clients appears at that base station.
Instead, Dist Est achieves roughly the same performance as Centralized, which however suffers from a slightly higher number of dissatisfied users because of the inconvenience of forcing all the transactions to pass through the root node, i.e., the core network, which is more prominent at high loads.
Finally, Dist Est enjoys smaller delays than Dist Probe in almost all cases, but the performance gap lessens as the load increases.
In particular, the fraction of dissatisfied users for the two algorithms is the same, though Dist Probe incurs a much more exorbitant cost in terms of network consumption, as discussed below.

\myfigeps{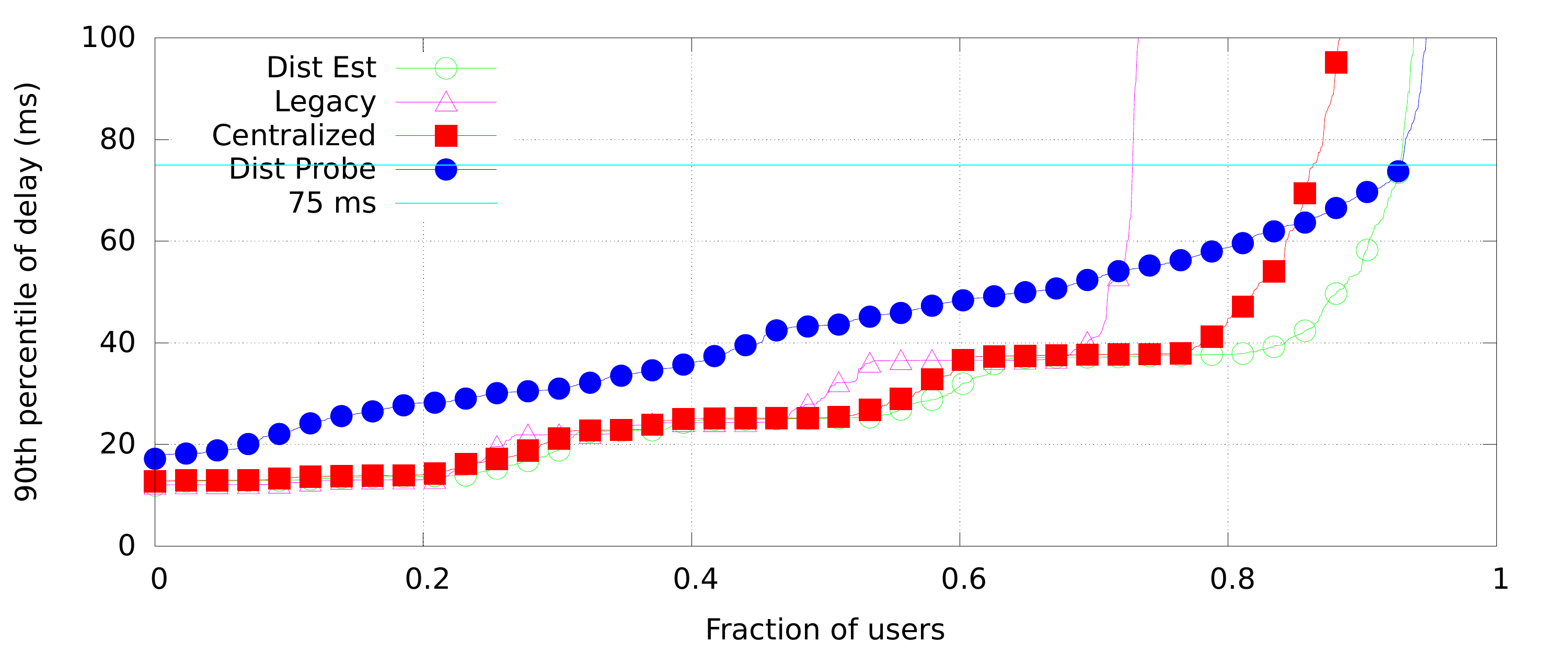}{Large-scale experiment: Distribution of the 90th of delay.}

This effect of excessive resort to network, which negatively affects the delay performance, is shown directly in \rfig{milano_tpt_mean}, which reports the overall average per-drop network throughput, sorted on the y-axis values.
Clearly, Legacy has minimum network consumption because the computation offloading traffic never leaves the base station.
However, as can be seen, the network overhead caused by Dist Est with respect to Legacy is negligible compared to that of both Centralized and Dist Probe.
This confirms that our proposed dispatching algorithm is able to achieve a good trade-off between wise utilization of the computational resources available and protocol/architecture complexity to achieve this goal, which ultimately benefits application latency in addition to traffic exchange.

\myfigeps{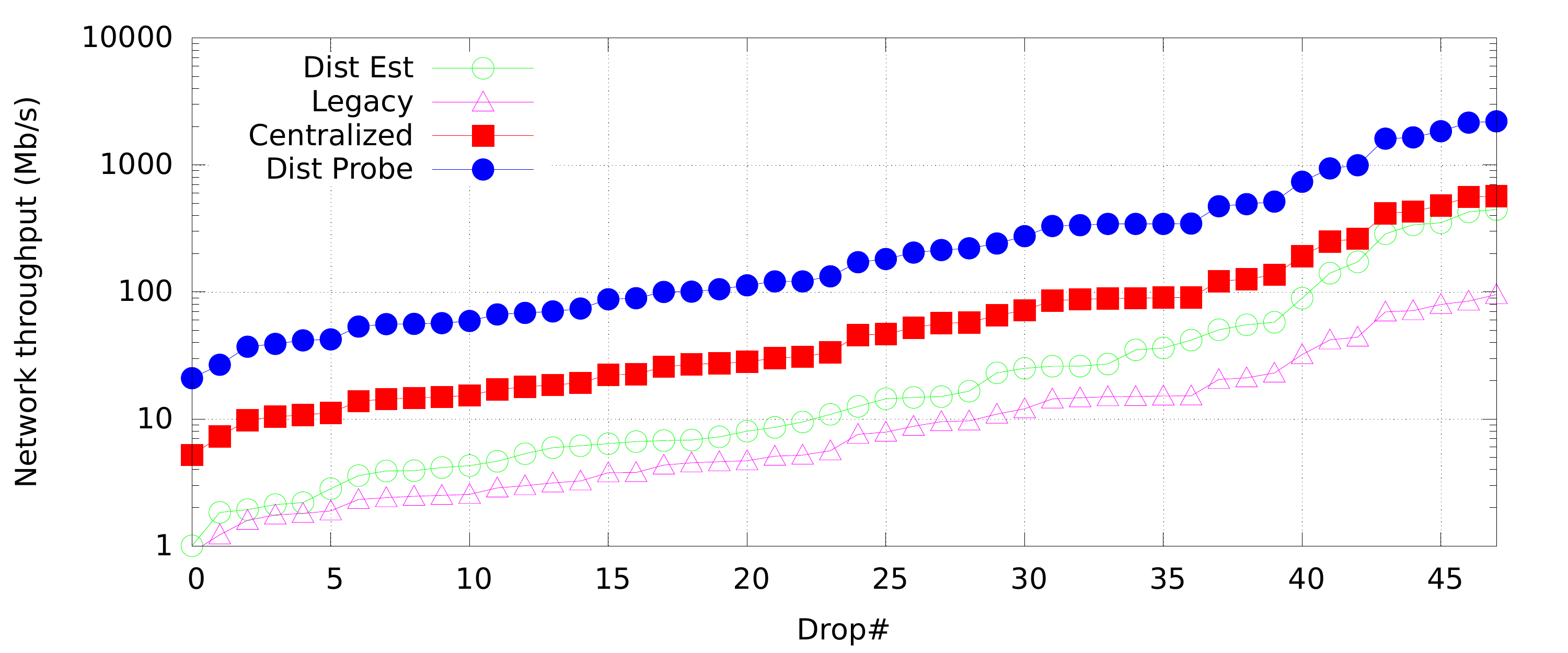}{Large-scale experiment: Network throughput per drop.}

\section{Conclusions}\label{sec:conclusions}

In this paper we have proposed a system to offload pervasive
applications with stringent delay requirements as stateless functions
on edge nodes with available computational capabilities, called
computers.
%
%
The execution of functions passes through edge nodes with dispatching
capabilities, which are ideally located as close as possible to the
final users.
The proposed architecture is highly scalable as the number of both
the clients and the computers grow
\added{%
because of its distributed
nature, in fact all edge nodes use only local information to dispatch
lambda functions, and thanks to the dispatchers only performing
passive measurements to estimate the communication and processing
times.

Furthermore, to overcome the limitations of existing frameworks for
the evaluation of edge computing systems, we have devised a novel
approach that jointly exploits network emulation, process-based
virtualization, and simulation of computation tasks to obtain
accurate results in controlled repeatable conditions with reasonable
computation effort.
We have used our framework to validate the distributed algorithm
to dispatch lambda functions in several scenarios.
}
The results have shown that the proposed distributed architecture
performs much better than statically allocating clients to computers
and the same as or better than both a centralized approach and a
distributed comparison solution from the literature.

\added{%
As future work, we will work on the integration of our performance
evaluation framework with edge computing platforms and communication
stacks to take into account in a realistic manner the overhead and
complexity introduced by, e.g., the \ac{ETSI} \ac{MEC} or the
\ac{LTE} \ac{EPC}.
}

\begin{acronym}
  \acro{ML}{Machine Learning}
  \acro{3GPP}{Third Generation Partnership Project}
  \acro{5G-PPP}{5G Public Private Partnership}
  \acro{AA}{Authentication and Authorization}
  \acro{AP}{Access Point}
  \acro{API}{Application Programming Interface}
  \acro{AR}{Augmented Reality}
  \acro{ARP}{Address Resolution Protocol}
  \acro{BGP}{Border Gateway Protocol}
  \acro{BS}{Base Station}
  \acro{CPU}{Central Processing Unit}
  \acro{EPC}{Evolved Packet Core}
  \acro{ETSI}{European Telecommunications Standards Institute}
  \acro{FCFS}{First Come First Serve}
  \acro{FSM}{Finite State Machine}
  \acro{GPU}{Graphics Processing Unit}
  \acro{HTTP}{Hyper-Text Transfer Protocol}
  \acro{ICN}{Information-Centric Networking}
  \acro{IoT}{Internet of Things}
  \acro{IIoT}{Industrial Internet of Things}
  \acro{ILP}{Integer Linear Programming}
  \acro{IP}{Internet Protocol}
  \acro{IPP}{Interrupted Poisson Process}
  \acro{ITS}{Intelligent Transportation System}
  \acro{ITU}{International Telecommunication Union}
  \acro{KPI}{Key Performance Indicator}
  \acro{LHS}{left-hand-side}
  \acro{LL}{Link Layer}
  \acro{LTE}{Long Term Evolution}
  \acro{MBWA}{Mobile Broadband Wireless Access}
  \acro{MCC}{Mobile Cloud Computing}
  \acro{MAC}{Medium Access Layer}
  \acro{MEC}{Multi-access Edge Computing}
  \acro{ML}{Machine Learning}
  \acro{MNO}{Mobile Network Operator}
  \acro{NAT}{Network Address Translation}
  \acro{NFV}{Network Function Virtualization}
  \acro{OF}{OpenFlow}
  \acro{OS}{Operating System}
  \acro{OSPF}{Open Shortest Path First}
  \acro{OWC}{OpenWhisk Controller}
  \acro{PMF}{Probability Mass Function}
  \acro{PU}{Processing Unit}
  \acro{QoE}{Quality of Experience}
  \acro{QoS}{Quality of Service}
  \acro{RHS}{right-hand-side}
  \acro{RPC}{Remote Procedure Call}
  \acro{RPC}{Remote Procedure Call}
  \acro{RR}{Round Robin}
  \acro{RSU}{Road Side Unit}
  \acro{TCP}{Transmission Control Protocol}
  \acro{TSN}{Time-Sensitive Networking}
  \acro{SDN}{Software Defined Networking}
  \acro{SMP}{Symmetric Multiprocessing}
  \acro{SRPT}{Shortest Remaining Processing Time}
  \acro{UDP}{User Datagram Protocol}
  \acro{URL}{Uniform Resource Locator}
  \acro{UT}{User Terminal}
  \acro{VANET}{Vehicular Ad-hoc Network}
  \acro{VM}{Virtual Machine}
  \acro{VNF}{Virtual Network Function}
  \acro{VPU}{Vision Processing Unit}
  \acro{WLAN}{Wireless Local Area Network}
  \acro{WRR}{Weighted Round Robin}
\end{acronym}

\end{document}